# Universal click-chemistry approach for the DNA functionalization of nanoparticles

*Nicole Siegel*[1], *Hiroaki Hasebe*[2], *Germán Chiarelli*[1], *Denis Garoli*[3,4], *Hiroshi Sugimoto*[2], *Minoru Fujii*[2], *Guillermo P. Acuna*[1,5]*, *Karol Kołątaj*[1,5]*

[1] Department of Physics, University of Fribourg, Chemin du Musée 3, Fribourg CH 1700, Switzerland.

[2] Department of Electrical and Electronic Engineering, Graduate School of Engineering, Kobe University, Kobe 657-8501, Japan.

[3] Dipartimento di Scienze e Metodi dell'Ingegneria, Università di Modena e Reggio Emilia, Via Amendola 2 Padiglione Tamburini, 42122 Reggio Emilia, Italy.

[4] Istituto Italiano di Tecnologia, Via Morego 30, 16163, Genova, Italy

[5] Swiss National Center for Competence in Research (NCCR) Bio-inspired Materials, University of Fribourg, Chemin des Verdiers 4, CH-1700 Fribourg, Switzerland.

*Corresponding authors.

Email address: guillermo.acuna@unifr.ch, karol.kolataj@unifr.ch.

**Abstract**

Nanotechnology has revolutionized the fabrication of hybrid species with tailored functionalities. A milestone in this field is the DNA conjugation of nanoparticles, introduced almost 30 years ago, which typically exploits the affinity between thiol groups and metallic surfaces. Over the last decades, developments in colloidal research have enabled the synthesis of an assortment of non-metallic structures, such as high-index dielectric nanoparticles, with unique properties not previously accessible with traditional metallic nanoparticles. However, to stabilize, integrate and provide further functionality to non-metallic nanoparticles, reliable techniques for their functionalization with DNA will be crucial. Here, we combine well-established dibenzylcyclooctyne-azide click-chemistry with a simple freeze-thaw method to achieve the functionalization of silica and silicon nanoparticles, which form exceptionally stable colloids with a high DNA surface density of 0.2 molecules/nm$^2$. Furthermore, we demonstrate that these functionalized colloids can be self-assembled into high-index dielectric optical antennas with a yield of up to 78% via the use of DNA origami. Finally, we extend this method to functionalize other important nanomaterials, including oxides, polymers, core-shell and metal nanostructures. Our results indicate that the method presented herein serves as a crucial complement to conventional thiol functionalization chemistry and thus greatly expands the toolbox of DNA-functionalized nanoparticles currently available.

## 1. Introduction

The field of colloidal nanotechnology has expanded significantly over the past few decades and currently covers a wide spectrum of shapes and materials, such as metal, dielectric, oxide, carbon, magnetic and polymeric nanoparticles (NPs).[1,2] Notably, these kinds of nanomaterials display unique optical, electronic, magnetic, mechanical, physical and chemical properties that are otherwise concealed in the bulk.[3–8] However, NPs are thermodynamically and energetically unfavorable relative to bulk materials, and even mild, physiological buffer conditions can induce aggregation.[9–11] Therefore, to take full advantage of their properties, it is typically necessary to modify the surface of NPs with stabilizing moieties.[12] Furthermore, NP functionalization also enables the fabrication of complex geometries and the combination with other species to provide additional tailored functionality.[13–20] Due to its unique specificity, addressability and biocompatibility, DNA is one of the most common choices for this purpose. Consequently, DNA-functionalized NPs have already found many applications in colloidal self-assembly, biosensor development, and drug delivery.[15,21–25]

The first chemical conjugation of NPs with DNA was introduced in 1996 with the landmark works of Mirkin and Alivisatos et al.[26,27] The method relies on a covalent binding of thiol-functionalized DNA strands to a metallic surface and is still one of the most widely used functionalization approaches. As both the surface of the Au NPs and the DNA strands are negatively charged, an effective means of repulsion screening is required for efficient functionalization.[28] Typically, this is realized by high salt concentration in a process known as 'salt aging'.[26,28,29] However, to avoid aggregation, salt has to be added very slowly, making the whole method highly time-consuming.[29] Recently, many new methods have been developed to reduce the functionalization times using either chemical (e.g., acids, surfactants, organic solvents) or physical means (e.g., microwaving and freezing).[29–33] Among these, freezing is particularly well-suited because it does not introduce any additional chemicals nor involves elevated temperatures, both of which could negatively affect the properties of the NPs.[33]

In general, these established methods provide a reliable way to functionalize only noble and coinage metals.[34–37] Therefore, it is necessary to develop alternative techniques to address the DNA conjugation of NPs with surface compositions that are inert to thiols, including oxides, core-shell, polymeric NPs, and even NPs composed of active metals (Al, Ga) and metalloids (Si, Ge). Importantly, these NPs exhibit various and unique physicochemical properties that give access to a broad range of applications from drug delivery to bio-sensing, bio-imaging and the manipulation of light at the nanoscale, as optical antennas.[38–44] Si NPs are an interesting example of structures that cannot be functionalized using thiol chemistry. These high-index dielectric NPs have received growing attention in recent years, as theoretical and numerical studies have predicted novel properties including low heat losses and overlapping strong magnetic and electric modes in the visible range.[44] These features are key for applications in chiral sensing, unidirectional scattering, and the manipulation of magnetic dipole transitions[45–50], making Si NPs a convenient alternative to plasmonic NPs that exhibit relatively high losses and the absence of a magnetic response at optical frequencies.[44] However, in spite of the great potential of Si NPs, their conjugation with DNA strands has yet to be achieved,[48,51–59] in part due to a lack of stable commercially available Si NPs. Moreover, it was only three years ago that the first colloidal suspensions of spherical Si NPs with resonances in the visible range, high yield and reduced size and shape dispersion were first introduced.[51,60,61]

Recently, a click chemistry reaction known as strain-promoted alkyne azide cycloaddition (SPAAC) has been increasingly explored as an alternative to thiols for DNA conjugation.[62–65] Not only does SPAAC benefit from very high specificity between dibenzylcyclooctyne (DBCO) and azide groups, but it also works in both aqueous and organic solutions and does not require elevated temperatures, extreme pH conditions or high salt concentrations.[66] Furthermore, the reaction is not material-specific and can be applied to any type of NPs, as long as azide groups are present on the surface.[66,67] Similar to thiol functionalization, repulsion between azide-bearing NPs and DBCO-DNA strands can diminish the final DNA surface density and lead to long reaction times, something that is particularly limiting for biological applications.[2] However, to the best of our knowledge, little progress has been made towards overcoming this limitation, with only one report on the use of the salt aging method to electrostatically screen the NP and DNA reactive groups.[65]

Inspired by recent developments in the functionalization of Au NPs, we report here an approach to conjugate various nanomaterials with DNA by employing freezing-assisted SPAAC, with a special focus on Si and $SiO_2$ NPs. Impressively, this simple method of functionalization proved to tremendously improve the overall DNA deposition efficiency. When comparing with other approaches used for the DNA functionalization with SPAAC, we found that not only does freezing reduce the reaction time from about two days to only a few hours, but also that our method yields 2-10 times higher DNA density on the surface of the NPs, i.e., up to 0.2 molecules/nm$^2$. The resultant DNA-Si NPs form highly stable colloids, which we incorporate into DNA origami structures via the formation of self-assembled NP dimers at pre-determined positions with an inter-particle gap of approximately 5-10 nm. Our results show up to a 78% yield for the formation of NP dimers (i.e., versus unintended NPs monomers), thus representing the first step towards the synthesis of more complex geometries such as hybrid metal-dielectric antennas and the combination of high-index materials with other species in precise geometries. Finally, we demonstrate the versatility of this method by realizing the DNA functionalization of different NPs: oxide – $TiO_2$, Au@$SiO_2$, polymeric – PMMA and PS, as well as metallic Au NPs.

## 2. Results and discussion

### 2.1. DNA functionalization of $SiO_2$ and Si NPs

$SiO_2$ NPs are well studied, widely available and easy to produce.[68] Moreover, the oxide-rich surface enables the use of silane chemistry to decorate the NPs with azide ($N_3$) for further DNA functionalization via SPAAC. Therefore, these particles constitute an excellent reference point for a thorough investigation of the combination of SPAAC with freezing processes to enhance the efficiency of the former. On the other hand, Si NPs feature desirable optical properties that stem in part from heat losses that are lower than those of Au NPs and to new phenomena based on the additional magnetic resonances that can lead to specific applications in light manipulation.[55,58] However, an efficient method for coating both types of NPs with DNA is still lacking. Given that Si NPs typically bear an intrinsic oxide layer, the results we present herein for $SiO_2$ NPs are expected to be easily translated to Si NPs.

To address this challenge, we devised a protocol for the DNA functionalization of $SiO_2$ and Si NPs by combining SPAAC click chemistry with a simple freezing protocol (**Figure 1**). First, the NPs are reacted with a chloride-terminated silane (CPTMS) in anhydrous DMF (**Figure 1a**), followed by reaction with sodium azide and ultimately the deposition of DBCO-bearing DNA on the azide-NPs via the formation of a stable triazole bond between DBCO and azide molecules (**Figure 1b**).[69] While such DBCO-azide click chemistry is a reliable method for functionalizing NPs, around two days are typically required to achieve sufficient grafting density.[62–64] Thus, to circumvent such a long deposition time, the suspension of azide-NPs and DBCO-DNA in PBS buffer was frozen at -20°C for 2 h in order to concentrate the DNA at the surface of the NPs via confinement in liquid pockets between ice crystals (**Figure 1a**). Subsequently, the nanoparticle suspension was thawed to room temperature via sonication for 5min to obtain a stable colloid.

To verify our results and to further optimize the functionalization process, an in-depth analysis was carried out for each reaction step. First, the success of the initial silanization reaction and subsequent chloride-to-azide conversion was confirmed colorimetrically for both $SiO_2$ and Si NPs through a reaction with DBCO-bearing reference dyes. Upon mixing the azide-NPs and DBCO-bearing dyes, the dyes covalently attached to the NPs via the formation of a triazole linkage as confirmed by a visible color change in the NPs (**Figure 2a, Figure S1a**). The resultant silane surface density was quantified for $SiO_2$ using **Equation 1** by measuring the normalized extinction intensity at 650 nm (i.e., at the absorption maximum of the employed Cy5 dye, **Figure S1a**) after the freezing-assisted reaction between the NPs and DBCO-modified Cy5 (**Materials and Methods**). To rule out non-specific binding to the NPs induced by freezing[70,71] and to verify the quantification method, the reaction was carried out with bare and chloride-terminated NPs as references. Obtained results confirm that freezing does not change the reaction specificity and that only the azide-functionalized nanostructures exhibit a color change after mixing with Cy5 and subsequent washing to remove any unreacted dye (**Figure S1a**). Furthermore, when up to ca. $10^4$/nm$^2$ CPTMS was used for the initial silanization reaction, the silane surface coverage obtained for NPs could be enhanced by increasing the amount of CPTMS added to the reaction (**Figure**

**S1b**). However, gains in surface coverage were diminished above $10^5$/nm$^2$ CPTMS, which we attribute to cross-bonding between the NPs as evidenced by $SiO_2$ and Si aggregates observed by DLS (**Figure S1c, Figure S2a**). Efforts to carry out such quantification for Si were impeded by the fact that the inherently high extinction coefficient of Si NPs conceals the absorption signal of Cy5.

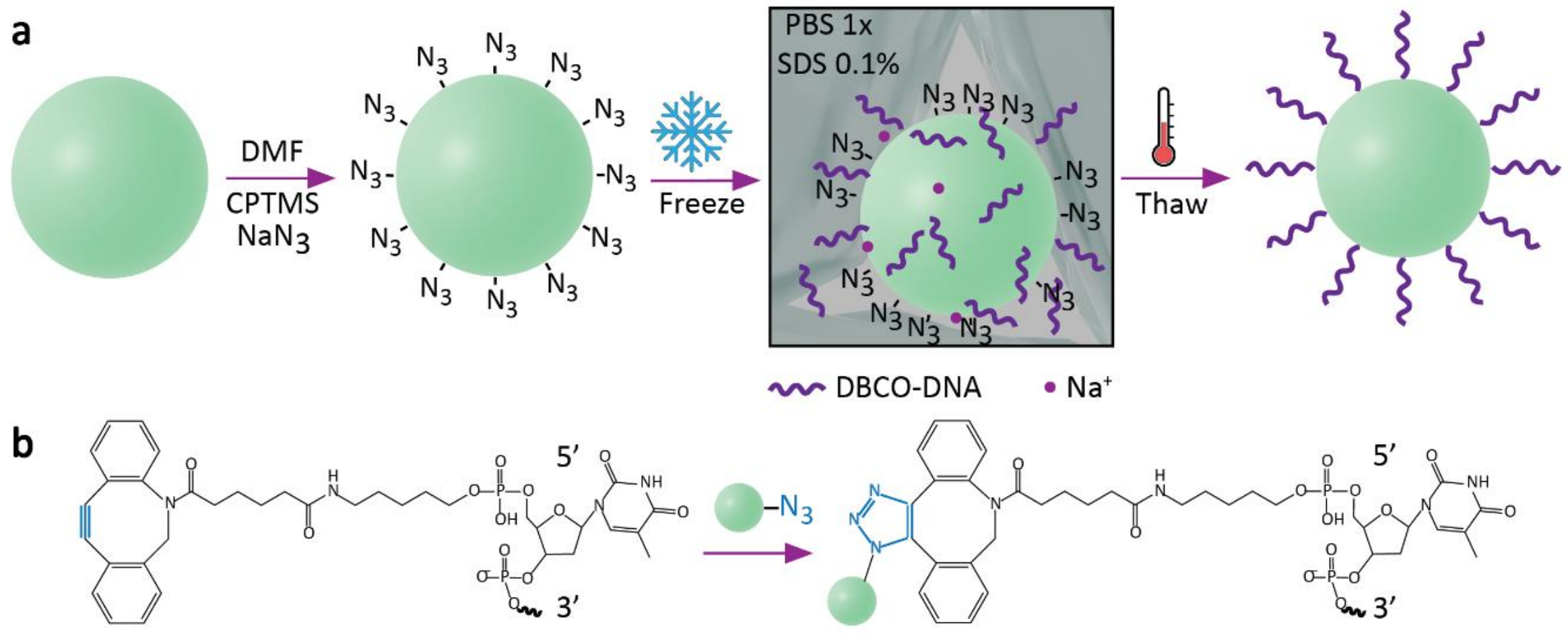

**Figure 1: The freezing-assisted SPAAC protocol.** (a) Schematic depiction of the two-step DNA grafting of oxide-rich NPs, consisting of azide functionalization via CPTMS silanization and chloro-azide substitution in anhydrous DMF, followed by freezing and subsequent thawing of the NPs in the presence of DBCO-DNA. (b) Depiction of the strain-promoted azide-alkyne cycloaddition (SPAAC) reaction employed in the functionalization protocol.

Considering the aforementioned unique features of DNA functionalized Si NPs, we studied the reaction of DBCO-bearing DNA strands with the azide-functionalized Si NPs. **Figure 2a** presents an image of thawed colloidal Si NPs after the freezing reaction with an excess of a DBCO-functionalized DNA sequence consisting of 22 thymidines (DBCO-T22). Notably, the bare NPs aggregate during the reaction and the colloid loses its color (**Figure 2a**, right panel), while the Si-$N_3$ NPs after DBCO-T22 conjugation remain stable. The observed difference in stability is corroborated by DLS measurements (**Figure S2a**) showing that the hydrodynamic diameter ($D_H$) of bare Si NPs increases significantly after freezing and thawing, while no substantial change in $D_H$ is observed for the azide-decorated structures after the freezing reaction with DBCO-T22. Moreover, no significant change in $D_H$ is observed for Si-T22 NPs between pH 2-10, in contrast to the ca. two-fold increase in $D_H$ observed for Si-$N_3$ NPs when decreasing the pH of the solution from 6 to 4 (**Figure S2b**). Zeta potential measurements further confirm successful DNA-functionalization, given that lower surface potentials are observed for Si-T22 NPs in comparison to unreacted Si-$N_3$ NPs (**Figure S2c**). It is worth noting that the lowest value of -37 eV was measured for the reaction ratio of CPTMS/surface of $10^2$-$10^3$/nm$^2$ (**Figure S2c**), and that the highest electrophoretic mobility was observed for Si-T22 NPs subjected to a CPTMS/surface ratio of $10^2$-$10^3$/nm$^2$, suggesting that this stoichiometry results in the most efficient DNA conjugation. Taken together, these data provide strong evidence that the successful functionalization of Si-$N_3$ NPs with DNA imparts stability to the NPs by screening the electrostatic interactions between the salt and the NPs that otherwise results in aggregation.[33]

Subsequently, we conducted electron energy loss spectroscopy (EELS) measurements of Si-T36 NPs ($d$ = 140 nm as calculated by their extinction spectra and measured by TEM) to investigate the morphology of the NPs after conjugation with DBCO-DNA (**Figure 2b, Figure S3**). This longer sequence comprised of 36 thymine bases was chosen for these experiments in order to obtain higher resolution and contrast. The TEM micrograph overlaid with the corresponding EELS elemental analysis measurements shows strong, overlapping oxygen and phosphorus signals along the perimeter of the NPs, indicating that significant amounts of DNA are present on the surface of the NPs (**Figure 2b, Figure S3c**). In contrast, no phosphorous signal was observed in neither TEM-EELS nor energy dispersive X-ray (EDX) control measurements carried out for both Si and Si-$N_3$ NPs (**Figure S3a,b**). Finally, cryo-EM measurements of Si-T36 NPs confirm the existence of a 10 nm-thick homogenous DNA shell on the surface of the NPs, which is consistent with the expected length for a T36 DNA strand and provides unequivocal proof of successful DNA-NP conjugation (**Figure 2c**, **Figure S4**).[72,73] Despite the clear success of this method, quantification of the density of DNA at the NP surface after functionalization was not possible, since, as with DBCO-Cy5, the Si extinction signal interferes with the detected absorption of the DNA. Therefore, to better assess the density of the DNA coating, we investigated DNA-functionalized $SiO_2$ nanostructures as a reference material. To this end, $SiO_2$ NPs were functionalized with DBCO-T22 in the same way as the Si NPs, and the density of DNA was quantified using **Equation 2** by extinction measurements at 260 nm (**Figure S5a**, see Materials and Methods for details). The results reveal surface densities between 0.05-0.2 DNA/nm$^2$, with the maximum surface density obtained for a reaction ratio of $10^4$ CPTMS/nm$^2$ (**Figure S5b**). The lower surface density of 0.2 molecules/nm$^2$ for $SiO_2$-T22 (**Figure S5b**) compared to 0.85 molecules/nm$^2$ for $SiO_2$-Cy5 (**Figure S1b**) can be explained by the repulsion of negatively charged DNA on the surface of $SiO_2$-T22 NPs, which reduces the overall efficiency of the SPAAC reaction. Nevertheless, this surface density appeared to be sufficient to maintain NP stability during gel electrophoresis, as seen in the inset in **Figure S5a**. To rule out non-specific DNA binding to the surface, which has been known to occur for Au NPs[70], we also tested DBCO-T22 deposition on bare $SiO_2$ and chloride-functionalized nanostructures ($SiO_2$-Cl) (**Figure S5a**). The absence of a DNA extinction peak at 260 nm for $SiO_2$ and $SiO_2$-Cl, in contrast to the strong extinction band observed for $SiO_2$-$N_3$ after reaction with DBCO-T22 at this wavelength, confirms that non-specific adsorption on the surface of NPs is alone not sufficient to obtain a stable DNA shell.

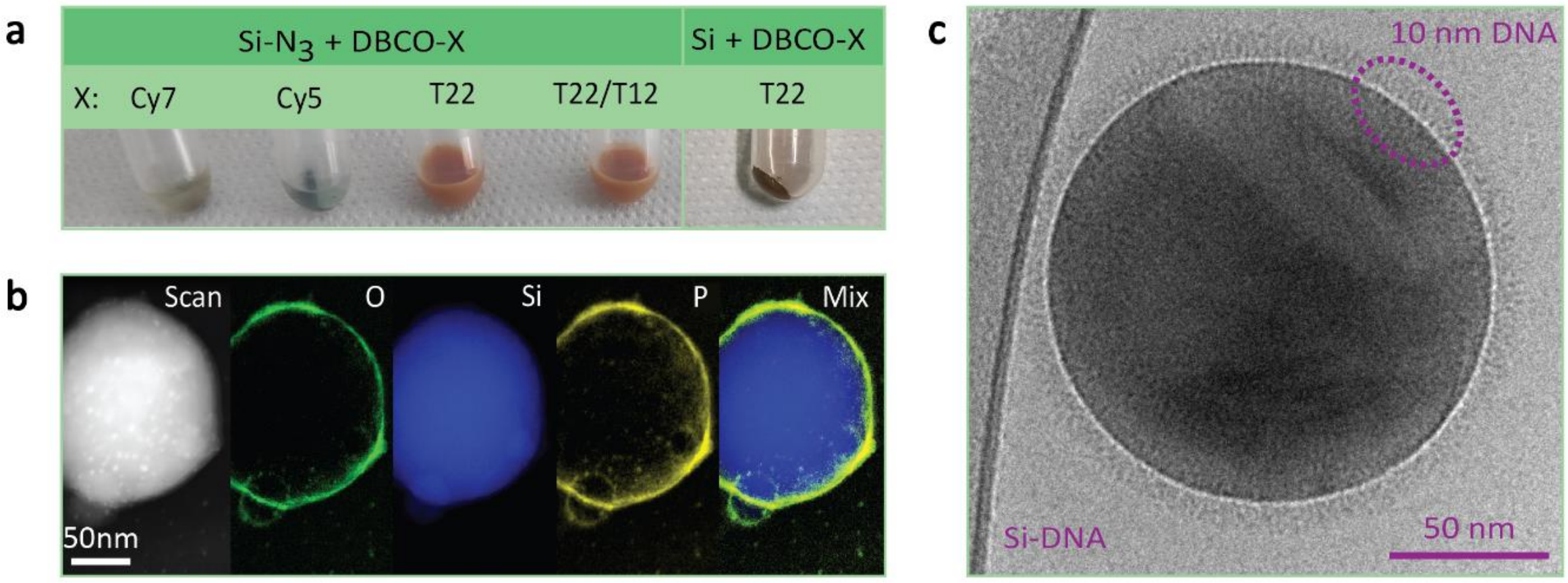

**Figure 2: DNA functionalization of Si NPs.** (a) Azide-functionalized Si NPs after mixing with DBCO modified with Cy5, Cy7, and DNA (either solely T22, or a 50:50 mixture of T12 and T22) compared with bare Si NPs mixed with DBCO-DNA. Bare Si NPs after thawing were spun down at 300 × g for 1 min for easier observation. (b) EELS elemental analysis of a Si NP (*d* = 140 nm) functionalized with T36 DNA strands showing (from left to right) the TEM scan, oxygen signal (green), silicon signal (blue), phosphorus signal (yellow), and their overlap (Mix). (c) Cryo-EM image of a DNA-functionalized Si NP (*d* = 140 nm). The T36 sequence that constitutes the DNA shell was measured to have an approximate length of 10 nm (purple circle).

To probe the versatility of our DNA-functionalization method, we sought to determine what, if any, influence the DNA sequence itself exerted on the DBCO-azide SPAAC reaction. Generally, the DNA functionalization of NPs cannot be considered as a straightforward reaction between two reactants (e.g., thiol-Au or DBCO-azide), as it is influenced by other parameters such as non-covalent interactions between the DNA sequence and the NP surface.[74–76] Such a sequence dependence has been demonstrated when conjugating thiol-DNA to Au NPs, with thymidine-rich sequences typically yielding significantly higher surface densities than other nucleotides.[75,77] To examine whether such a dependence exists with the functionalization approach presented herein, we carried out the freezing-assisted SPAAC reaction with $SiO_2$-NPs and either DBCO-modified poly-T (T18), poly-A (A18), poly-C (C18), poly-G (G18), or a random sequence CTTCTCTACCACCTACAT (R18). The amount of DNA conjugated to $SiO_2$ NPs was then calculated quantitatively using the previously described spectral method (**Figure S6a**). The results indicate that indeed the efficiency of the freezing-assisted SPAAC reaction strongly depends on the sequence employed, with the highest surface density obtained for G18 (ca. 0.3 molecules/nm$^2$), followed by T18, R18, A18 and C18 (ca. 0.05 molecules/nm$^2$) in descending order (**Figure S6a**). For all sequences other than C18, a significantly higher surface density of DNA was obtained with freezing-assisted SPAAC in comparison to non-azide-bearing control particles, suggesting that non-specific absorption of the C18 sequence hinders the functionalization reaction. This effect can be related to a high affinity of the cytosine residue to the NP surface, resulting in wrapping of the NP with this particular DNA sequence. This process has been already observed for adenine and cytosine-rich sequences for Au and Ag NPs, respectively.[70,78] DLS measurements carried out on both DNA-functionalized $SiO_2$ and Si NPs show similar trends in $D_H$ as a function of the DNA sequence employed (**Figure 3b and Figure S6b**), which are in line with the trend in sequence-dependent surface density as measured spectroscopically for $SiO_2$ (**Figure S6a**). Therefore, both $SiO_2$ and Si NPs with higher surface density show increased stability at high salt concentration, with the exception of the G18-decorated NPs that exhibit significant aggregation at salt concentrations as low as 5 mM $MgCl_2$. We hypothesize that the disproportionately low stability of the G18-functionalized NPs may be related to the formation of G-quadruplexes between individual NPs, a phenomenon that has been observed for Au NPs.[79] Finally, DNA-bearing Si NPs that exhibited lower $D_H$ (i.e., NPs for which less aggregation is observed) also showed greater mobility when subjected to gel electrophoresis (**Figure 3b,c**), further corroborating the relationship between particle $D_H$ and surface charge indicative of successful DNA functionalization.

Finally, an assessment of our freezing-assisted SPAAC method in comparison to alternative reported methods for functionalization of non-metallic particles, such as room-temperature mixing and salt aging,[62,65] shows that the approach reported herein more reliably yields NPs that retain their stability under harsh salt and pH conditions. First, our comparative measurements of DNA (T22) functionalization of $SiO_2$ NPs indicate that freezing substantially increases the amount of conjugated DNA from 0.01 and 0.07 DNA/nm$^2$ for the mixing and salt aging method respectively to approximately 0.13-0.16 DNA/nm$^2$ for the freeze-thaw process (**Figure S7a,b**). The high DNA surface density for the $SiO_2$-T22 NPs made via freezing-assisted SPAAC endows these particles with greater stability in the presence of 20 mM $MgCl_2$ than those made via mixing or salt aging (**Figure S7c**). A similar comparative study was carried out for Si NPs, which are generally less stable than $SiO_2$ NPs and readily aggregate at high salt concentration, as evidenced by a loss of color during salt aging (**Figure 3d**, middle row). In contrast to the salt aging process, no such aggregation is observed when freezing-assisted SPAAC is employed (**Figure 3d**, lower row), which we attribute to a fast formation of liquid pockets during the phase transition that persist after freezing and shield the NPs from long exposure to high salt concentrations. Moreover, after thawing, Si NPs functionalized via freezing-assisted SPAAC maintained their colloidal stability even at pH 12 or in the presence of 1 M NaCl, conditions under which Si-T22 NPs prepared via mixing or salt aging readily aggregate after only 10 min (**Figure 3d**). This difference in stability was also readily observed via gel electrophoresis, in which only the particles prepared via freezing-assisted SPAAC featured sufficiently high surface density of DNA to allow them to run on the gel (**Figure 3e**). Finally, a simple agarose gel electrophoresis after the sucrose-gradient separation (see **Materials and Methods**) was sufficient to decrease the size dispersion of Si NPs from 15% to 6% (**Figure 3e,f** and **Figure S8**). Considering that many of the optical properties of NPs, such as their scattering modes, depend on their size,[3,51] we see the possibility to fractionate nanostructures by gel electrophoresis as an important advantage of the freezing-assisted SPAAC, which is not offered by alternative functionalization methods.

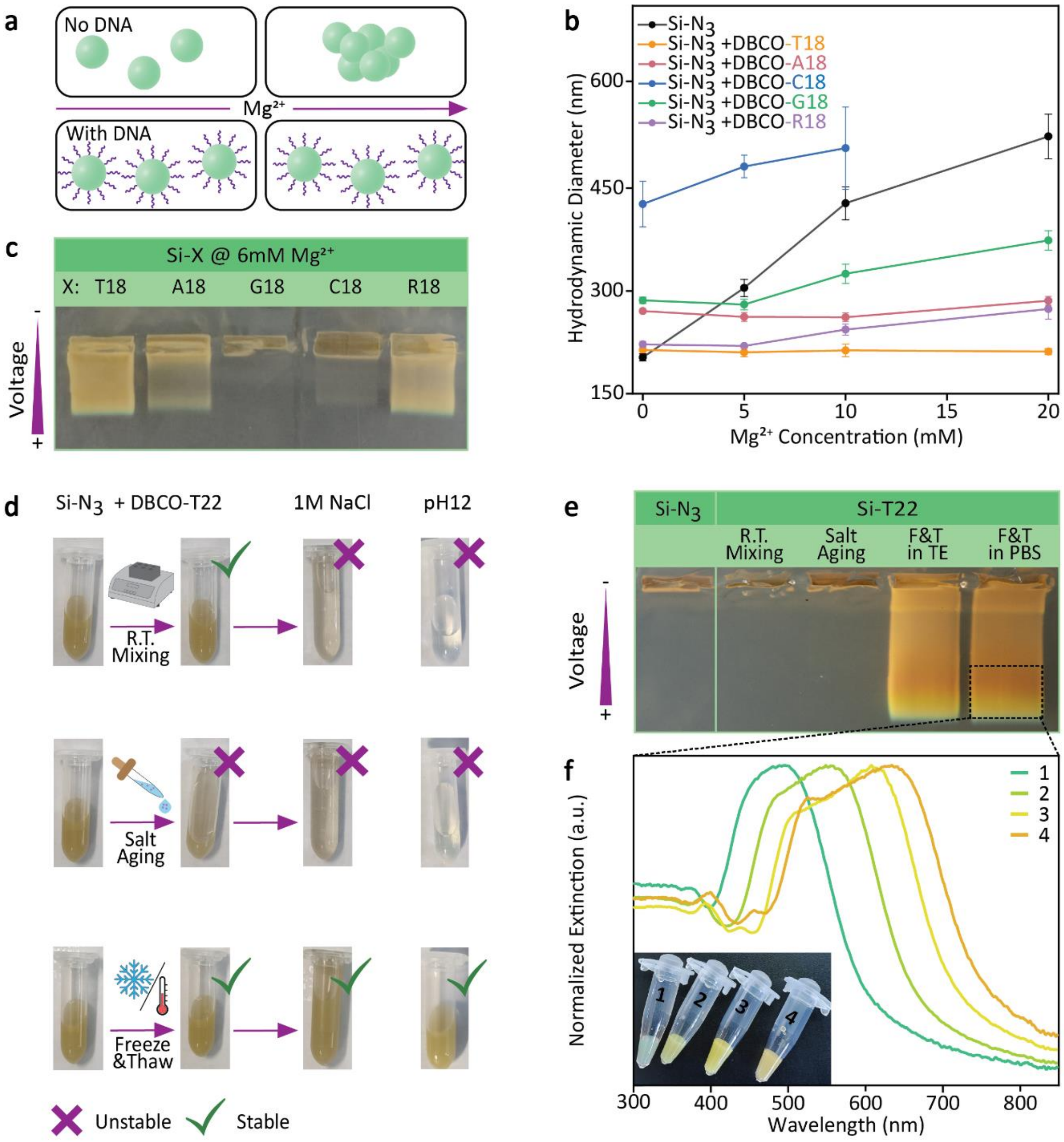

**Figure 3: Reaction characterization.** (a) Schematic depicting the tendency of bare nanoparticles to aggregate in the presence of magnesium salt (upper panels), as well as the resistance to such aggregation conferred to these particles via DNA-functionalization (lower panels). (b) Hydrodynamic diameter of Si NPs as a function of salt concentration. Si NPs were functionalized with either $N_3$ (black), $N_3$ + DBCO-T18 (orange), $N_3$ + DBCO-A18 (pink), $N_3$ + DBCO-C18 (blue), $N_3$ + DBCO-G18 (green) and $N_3$

+ DBCO-R18 (violet). No data is included for C18-Si NPs above 15mM $Mg^{2+}$ due to the fact that these particles formed extremely large aggregates. Error bars represent one standard deviation, and solid lines serve as a guide to the eye. (c) Image showing the result of a gel electrophoresis carried out at 6 mM $Mg^{2+}$ for the DNA-functionalized Si NPs. (d) Images of suspensions of azide-decorated Si NPs in phosphate buffer saline before and after functionalization with DBCO-DNA using room-temperature mixing for 48 hours (upper row), salt aging (SA, middle row), or freezing-assisted SPAAC (F&T), together with images 10 min after the addition of 1M NaCl, or after the addition of KOH until pH 12 was reached. In each case, particle stability (instability) is indicated by a check mark (cross). (e) Image showing the result of a gel electrophoresis run of Si-$N_3$ NPs and DNA-functionalized Si NPs with different methods: room-temperature mixing, salt aging, and freezing-assisted SPAAC both in Tris-EDTA buffer (TE) and phosphate-buffered saline (PBS). (f) Normalized extinction spectra of Si-T22 NPs obtained after size-separation via gel electrophoresis. Inset: image showing each solution after removal from the gel. All electrophoretic gels consisted of 0.3% agarose in 0.5x TEA and 6 mM $Mg^{2+}$.

### 2.2. Si NP optical antenna assembly with DNA Origami

Over the last decade, the DNA origami method[80] has established itself as one of the most versatile tools for the bottom-up synthesis of hybrid species with tailored functionality.[81,82] This approach, which is based on the self-assembly of DNA strands into rationally designed 3D structures, has been exploited to organize different entities, such as fluorophores, proteins, or NPs with nanometer precision and stoichiometric control.[83,84] Consequently, DNA origami has been widely employed to produce optical antennas with controlled geometries for sensing and light manipulation at the single-molecule level.[85,86,86–91] However, the incorporation of NPs into DNA origami typically must be carried out in salt buffers, thus requiring that the NPs retain their colloidal stability in high-salt environments. To date, NP manipulation with DNA origami has therefore been mostly limited to Au, Ag and quantum dot NPs.[81] To the best of our knowledge, no studies reported so far on the incorporation of high-index dielectric NPs in DNA origami scaffolds exhibiting low heat losses together with electric and magnetic resonances in the visible range. As a result, little is known about the interaction of Si NPs with light at the single-molecule level in control geometries, which is crucial for the validation of current numerical predictions on their optical properties and to explore new applications for high-index dielectric nanomaterials in fields ranging from nanophotonics to chiral sensing.

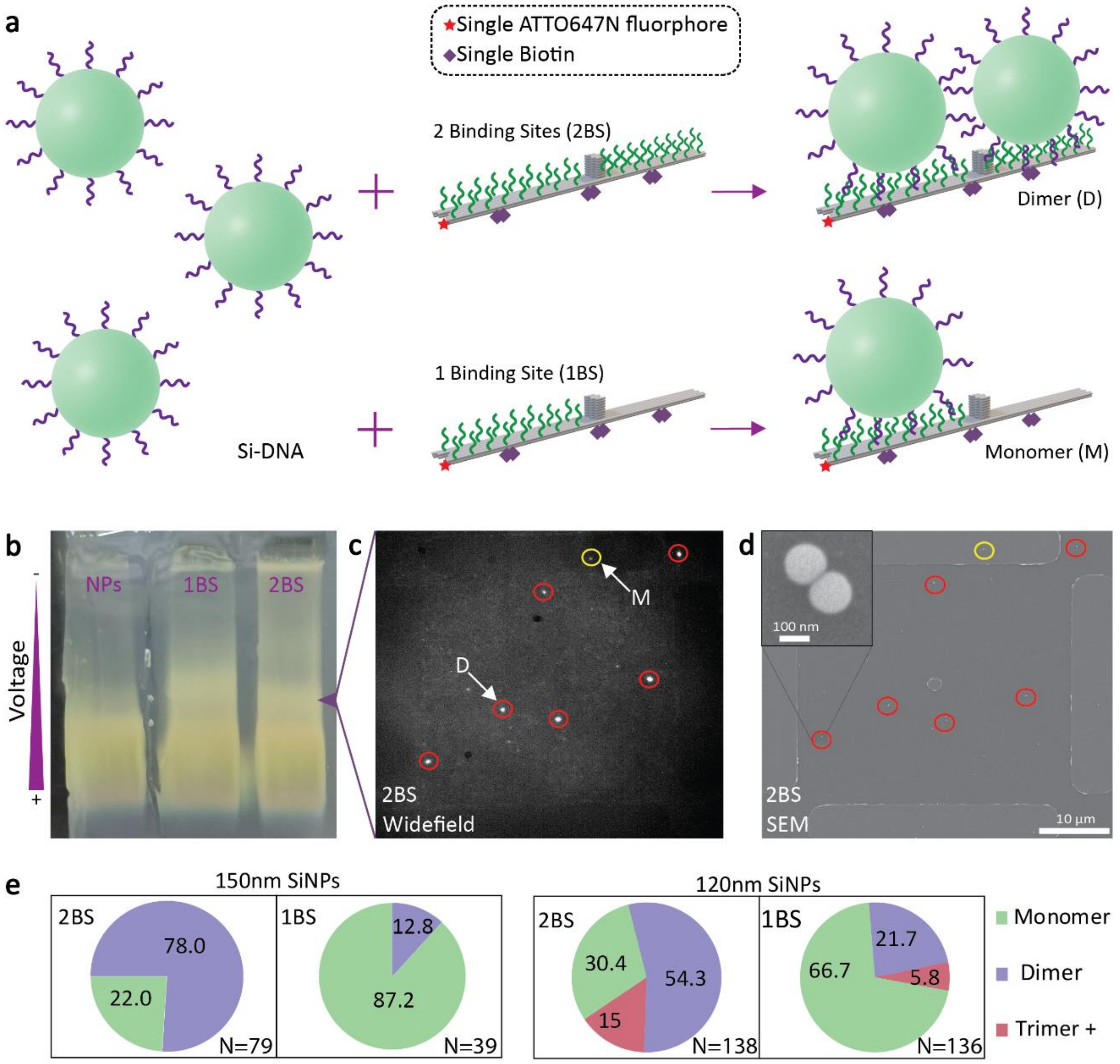

**Figure 4: Si NP assembly onto DNA origami.** (a) Schematic depicting the assembly of DNA-functionalized Si NPs onto DNA origami structures. The DNA origami contains two rows of poly-adenine (green protrusions) on one or both sides of the mast for the formation of Si monomers (1BS) or dimers (2BS), a single ATTO 647N fluorophore (red star), and six biotin molecules (violet diamonds) for the binding to the surface of a glass slide. (b) Agarose electrophoresis gel through which the dimer design structure (2BS), monomer design structure (1BS) and NPs were run, with the band of interest from the 2BS lane indicated with an arrow. (c) WF and (d) SEM colocalization of the 2BS origami design showing dimers (red) and monomers (yellow). The inset in (d) shows the magnification of one of the dimers. (e) Statistics carried out on the WF-SEM colocalization results depicted in (c,d). with the percentage of monomers in green, dimers in violet and higher-order structures in pink for both 2BS and 1BS origami designs after assembly with 150 nm and 100 nm DNA-functionalized Si NPs.

Thus, we sought to exploit the exquisite stability of the DNA-bearing Si NPs fabricated by our freezing-assisted SPAAC method to assemble Si NP dimers using custom-made DNA origami scaffolds. To this end, we used a high aspect ratio framework of DNA origami comprised of a 12-helix bundle with a final, folded length of 180 nm and a 20 nm mast in the middle (**Figure S9a**). To facilitate the binding of the NPs to the scaffold, the DNA origami was equipped with either one (1BS) or two binding sites (2BS), with each site featuring 16 adenine handles to dock a single particle. As conceptually presented in **Figure 4a** and **Figure S9**, this approach allows for the fabrication of either monomer or dimer structures on the 1BS and 2BS scaffolds, respectively. Additionally, a single DNA strand of each origami structure was modified with an ATTO 647N fluorophore for surface detection using wide-field (WF) microscopy, and six DNA strands were labelled with biotin molecules for biotin-neutravidin binding to the surface of the glass slide. Synthesis of Si monomer and dimer structures was achieved by following established protocols for the assembly of DNA-functionalized Au and Ag NPs onto DNA origami.[81] To this end, Si NPs were first functionalized with a 1:1 ratio of T18/28 DNA strands using freezing-assisted SPAAC. This mixture of long (T28) and short (T18) strands was employed with the goal of obtaining a DNA shell with well-separated T28 strands, which is expected to better facilitate hybridization with the complementary DNA strands on the origami scaffold. Next, the DNA-functionalized NPs were separated from excess, unreacted DNA strands by gel electrophoresis (**Figure 4b**). The purified NPs were subsequently incubated with the DNA origami structures overnight at room temperature to obtain either monomers or dimers of Si NPs as schematically depicted in **Figure 4a**. Purification of the origami-NP assemblies (i.e., 1BS and 2BS) by gel electrophoresis revealed the appearance of additional electrophoresis bands that indicate successful attachment of the NPs to the origami scaffolds, with some smearing of the bands resulting from variation in particle size (**Figure 4b**). Crucially, the positions of the Si NPs as determined via scanning electron microscopy (SEM) were found to colocalize with the positions of the origami-bound ATTO 647N molecules as detected by fluorescence WF microscopy, thus providing unequivocal proof of successful origami-NP binding (**Figure 4c,d**). Statistics obtained from WF-SEM colocalization analysis show that, for the 1BS design combined with either 150 and 120 nm Si NPs, 87% (13%) and 67% (28%) of the localized nanostructures were monomers (dimers), respectively. On the other hand, when the 2BS DNA origami was used, samples incubated with 150 and 120 nm Si NPs featured 78% (22%) and 54% (30%) dimeric (monomeric) structures, respectively. It is worth mentioning that a small fraction of the detected dimers might arise from self-dimerization of DNA-functionalized NPs prior to incubation with DNA origami. Based on SEM measurements, we estimate this effect to account for maximum 10-20% of all detected dimers, and to at least partly explain the relatively high number of dimers observed for 1BS origami-NP assemblies (**Figure 4e, Figure S9b**). Taken together, these results confirm that positional control can be effectively exerted over the assembly of DNA-functionalized Si NPs using DNA origami, while further optimization of factors such as particle size distribution and incubation conditions could result in even higher purity of the desired monomeric or dimeric NP assemblies.

### 2.3. Versatility of the freezing-assisted SPAAC method for DNA deposition

The results obtained for the DNA functionalization of Si and $SiO_2$ NPs clearly show that the use of freezing-assisted SPAAC significantly increases the efficiency of NP surface reactions and, in doing so, allows to obtain DNA-bearing NPs that exhibit exceptional stability at high salt concentrations. To further explore the versatility of this method, we sought to employ freezing-assisted SPAAC to functionalize a range of distinct NPs, namely Au@$SiO_2$ (core-shell), $TiO_2$, polymeric and finally also metallic NPs. Au@$SiO_2$ NPs constitute a particularly interesting example of nanostructures with a hydroxyl-rich surface, as the outer silica shell protects the metallic Au core from the external environment and, therefore, preserves the plasmonic resonance of the core (**Figure 5a**).[40,92,93] However, the silica shell is impervious to the thiol surface chemistry typically employed to functionalize Au NPs with DNA,[2] as is the case for other oxide NPs such as $TiO_2$ that are widely exploited for, e.g., plasmonic catalysis.[94] In contrast to the difficulties encountered with thiol chemistry, we were able to readily functionalize the shell of Au@$SiO_2$ NPs, as well as the surface of $TiO_2$ NPs, with poly-T DNA sequences using our freezing-assisted SPAAC protocol. Successful functionalization with either DBCO-bearing reference dyes or (optionally fluorophore-tagged) poly-T sequences was confirmed for Au@$SiO_2$ and $TiO_2$ NPs via DLS measurements **(Figure S10a,c**), as well as with gel electrophoresis for Au@$SiO_2$ NPs (**Figure 5a**) and with a colorimetric assay for $TiO_2$ (**Figure 5b** and **Figure S10b**). Notably, these functionalized oxide particles exhibit the same stability at high salt concentrations as the DNA-bearing $SiO_2$ and Si NPs (**Figure S10a,c**), and thus their incorporation into DNA origami is expected to be straightforward.

Beyond oxide particles, polymeric particles are important for myriad applications such as drug delivery or bio-imaging.[41,42] Thus, we sought to investigate the effectiveness of our freezing-assisted SPAAC method for conjugating DNA to the surfaces of two types of commercially available azide-functionalized polymeric particles, namely PMMA-$N_3$ ($d$ = 2 μm) and PS-$N_3$ ($d$ = 5 μm) microspheres. Given that the amorphous character of these particles can often lead to physisorption of various molecules, we assessed the binding of DNA to their surfaces by carrying out freezing-assisted SPAAC using both DBCO-bearing and DBCO-free poly-R DNA sequences modified with a Cy3 dye (i.e., DBCO-R18-Cy3 and R18-Cy3, respectively). Indeed, some physisorption of R18-Cy3 is observed for both particle types after the freezing step, as indicated by the visual appearance and emission spectra of the polymeric NPs after the reaction (**Figure 5c** and **Figure S10d,e**). In contrast, the reaction with DBCO-R18-Cy3 yields 12 and 6.7 times higher grafting density for PMMA and PS particles, respectively (**Figure S10d,e**). Finally, we showed that Au NPs ($d$ = 60 nm) could also be functionalized by first decorating these particles with azide groups via reaction with SH-PEG5-$N_3$, followed by freezing-mediated functionalization with DBCO-T18 (**Figure 5d** and **Figure S10f**). Thus, not only does our methodology enable the conjugation of DNA to a diverse array of nanoparticles not previously accessible with conventional functionalization methods but can also be used as an alternative to the well-established thiol-chemistry typically employed to functionalize Au NPs.

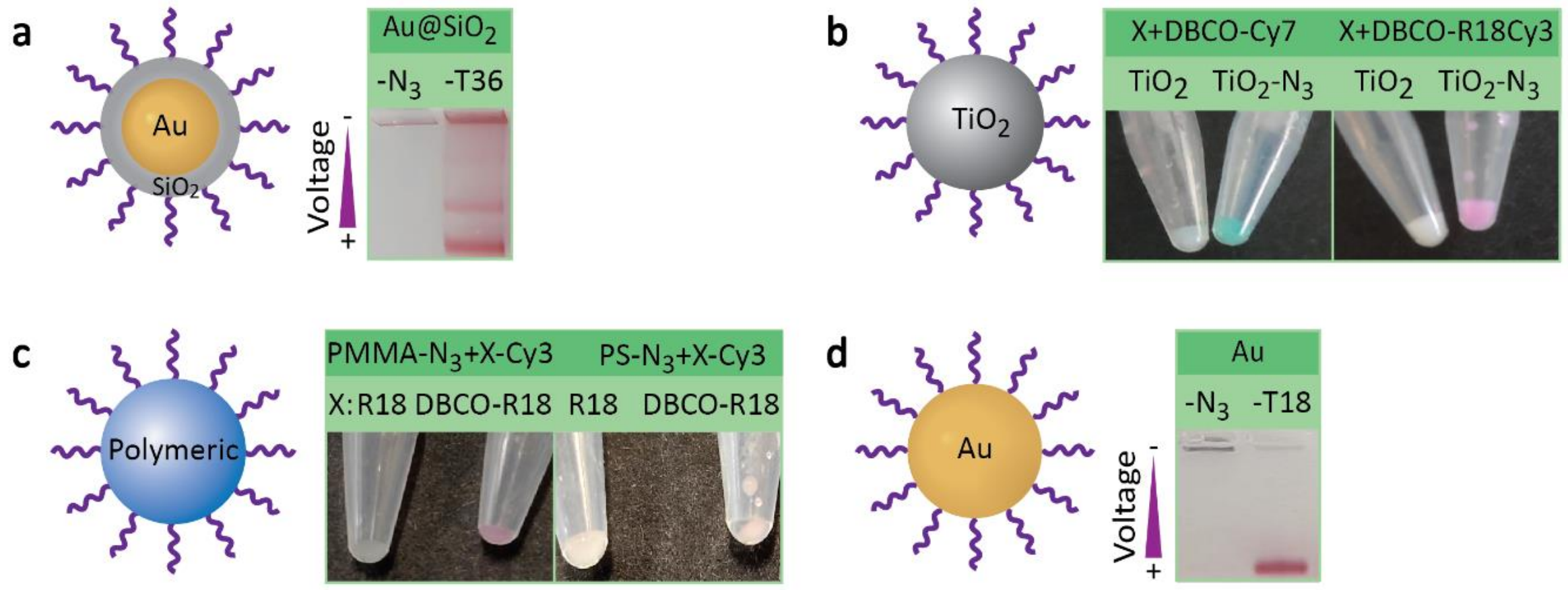

**Figure 5: Versatility of the freezing-assisted SPAAC method.** (a) Agarose gel electrophoresis of 60 nm Au@$SiO_2$ NPs before ($N_3$) and after (T36) DNA functionalization. (b) Images of $TiO_2$ (left) and $TiO_2$-$N_3$ (right) after reaction with DBCO-Cy7 (left panel) and with DBCO-R18-Cy3 (right panel). (c) Images of commercially available azide-functionalized 2 μm PMMA (left panel) and 5 μm PS (right panel) NPs after reaction with DBCO-R18-Cy3 and R18-Cy3. (d) Agarose gel electrophoresis of 60 nm Au@SH-$PEG_5$-$N_3$ before ($N_3$) and after (T18) DNA functionalization. Electrophoretic gels consisted of 1% agarose in 1x TEA and 12 mM $Mg^{2+}$.

## 3. Conclusions

In conclusion, in this work we report a simple and versatile technique for the DNA functionalization of various types of nanoparticles. In particular, we implemented the freeze-thaw method in combination with the DBCO-azide click chemistry to conjugate DNA to the surfaces of a variety of NPs of different materials. We initially focused on non-metallic NPs, in particular $SiO_2$ NPs due to their availability and ease of synthesis, as well as Si NPs due to their unique optical properties such as overlapping magnetic and electric resonances, together with low heat losses. Our results show that this combination of methods can be successfully applied for the DNA functionalization of these NPs. Furthermore, we have demonstrated that this method not only shortens the reaction time from about two days for previously reported approaches to an hour, but also increases the density of the DNA coverage by more than an order of magnitude. Thus, compared to other methods, the obtained NPs show excellent colloidal stability, a prerequisite for exploiting these NPs for new applications in various fields. As a proof of concept, we realized the assembly of high-index dielectric optical antenna dimers by incorporating Si NPs into DNA origami structures. Our measurements indicate that over 75% yield of the dimers can be obtained following the same simple incubation protocols established for incorporating Au and Ag NPs into DNA origami. Therefore, our results represent an important first step towards combining high-index dielectric nanoparticles and other species such as photon emitters at the single molecule level with nanometer precision and stoichiometric control. Finally, we demonstrate the versatility of our technique by functionalizing different NPs including oxides ($TiO_2$), core-shell particles (Au@$SiO_2$), polymers (PMMA, PS) and metals (Au) with DNA. These findings support the conclusion that the freezing-assisted SPAAC method reported herein constitutes a universal approach for the functionalization of essentially any type of nanoparticle and will thus promote a wide range of nanotechnological advancements in fields such as biosensing, drug delivery and self-assembly.

## 4. Materials and Methods

**Materials.** Methanol (MeOH, ≥99%), ethanol (EtOH, ≥99% and 99.8%), extra dry AcroSeal™ dimethylformamide (DMF, ≥99.9%), and neutravidin were purchased from Thermo Scientific. 3-chloropropyltrimetoxysilane (CPTMS, 97%), phosphate buffered saline pH 7.5 10x (PBS), tris acetate EDTA buffer 50x pH 8.0 (TAE), magnesium chloride ($MgCl_2$, 1M in $H_2O$), and DBCO-Cy7 were purchased from Alfa Aesar. Dimethylamine (DMA, 40% in $H_2O$), ammonia solution ($NH_4OH$, 28–30%), sodium azide ($NaN_3$), sodium dodecyl sulfate (SDS, 10% in $H_2O$), tetraethyl orthosilicate (TEOS, ≥99.9%), albumin biotin labeled bovine, 16-mercaptohexadecanoic acid (MHA), and DBCO-Cy5 were purchased from Sigma Aldrich. Thiol-PEG5-azido (SH-$PEG_5$-$N_3$) was purchased from Biosynth.

**Silica.** Silica NPs ($SiO_2$ NPs) with a diameter of 150 nm were prepared by the Ströber method.[68] Briefly, ethanol (285 mL), water (40 mL), and $NH_4OH$ (6.4 mL) were added to a 500 mL flask and stirred at 30 °C for 20 min. Afterwards, absolute TEOS (12 mL, 54 nmol) was added rapidly to the mixture and reacted for 3 h under stirring. Obtained NPs were purified by centrifugation at 5000 × *g* for 5 min and redispersed twice in methanol and twice in water consecutively. Finally, $SiO_2$ NPs were dried overnight on a desiccator before further functionalization with azide.

**Silicon.** Crystalline silicon NPs (Si NPs) were prepared by thermal disproportionation (1450 °C) of commercially available silicon monoxide (FUJIFILM WAKO) in $N_2$ atmosphere for 30 min. Si NPs were liberated from silicon dioxide matrices by etching in HF solution (46%) for 1 h, transferred to methanol and ultrasonicated for 1 min. The NPs were size separated by a sucrose gradient centrifugation process. First, sucrose solutions with different concentration (35-45 wt%) were added to a centrifuge tube from higher to lower concentration. Then, the Si NP solution was carefully deposited onto the sucrose solution and centrifuged at 4000 × *g* for 90 min. The size-separated Si NP solutions were retrieved from the top to the bottom of the solution and transferred to water.[51]

**Silica-coated gold.** The synthesis of Au@$SiO_2$ NPs was realized using the method developed by Xue et al.[95] Briefly, an ethanolic solution of MHA was added to commercially available 60 nm AuNPs (BBI) to the final concentration of 20 μM. This solution was then centrifuged at 5000 × *g* for 10 min and redispersed in an ethanolic solution of TEOS (0.1M). Finally, DMA (40%) was added while stirring until the concentration reached 0.6 M. After 3 h of the $SiO_2$ deposition, the obtained core–shell NPs were centrifuged twice for 10 min at 3000 × *g* and redispersed in ethanol. Finally, the NPs were dried overnight on a desiccator before further functionalization with azide.

**Titania.** Commercially available $TiO_2$ NPs with a diameter of 150 nm (Sigma Aldrich Prod. N: 914835) were dried overnight on a desiccator before further functionalization with azide.

**Azide-modified, oxide-rich NPs.** 2 pmol of dried Si, $SiO_2$, $TiO_2$ or Au@$SiO_2$ NPs were dispersed in anhydrous DMF (10 mL) by ultrasonication followed by addition of CPTMS (4 μl 97%, 21.6 μmol). After 2 h of sonication at 70 °C, $NaN_3$ was incorporated in excess and the mixture was stirred overnight at 40 °C. The NPs were purified by centrifugation at 5000 × *g* for 5 min and redispersion, twice in methanol and twice in water consecutively.[96]

**Azide-modified polymeric NPs.** 2 μM azide-modified PMMA and 5 μM azide-modified PS particles were purchase from PolyAn Molecular Surface Engineering (Prod. N: 108 35 002) and Spherotech (Prod. N: ACCP-50-5) respectively.

**Azide-modified gold NPs.** 60 nm AuNPs (OD1.5, 500 μL) were mixed with SH-PEG5-$N_3$ (14 μL 200 μM) and vortexed for 5 seconds. Afterwards 10% SDS (10 μL) was added, and the mixture was sonicated using a finger sonicator. The particles were then purified via centrifugation at 4000 × *g* for 5 min and resuspended in 0.1% SDS.

**DNA functionalization of NPs.** 1 mL of 400 pM NPs were centrifuged (time and speed were adapted for each NP) and resuspended in 100 µM DBCO-modified DNA (Biomers GmbH) to a final ratio of $10^3$ DNA/$nm^2$. Afterwards, 1 mL of the PBS-SDS buffer (1X PBS pH 7.5, 0.1% SDS) was added to the mixture, which was then frozen at -20 °C for 2 h. The NPs were thawed by sonication for 5 min before being centrifuged twice for 10 min at 2000 × $g$ and resuspended in water. For further purification and size separation a 0.3% agarose (LE Agarose, Biozym Scientific GmbH) electrophoresis gel in 0.5x TAE (20 mM Tris, 5 mM Acetate, 0.5 mM EDTA) and 6 mM $MgCl_2$ was run for 3 h at 100 V for the Si and $SiO_2$ particles. The NPs were recovered by squeezing the desired band from the gel.

**Particle analysis.** An Anton Paar (Litesizer 500) Z-sizer device was used for dynamic light scattering (DLS) and electrophoretic light scattering (ELS) measurements. For all measurements, 5 repetitions (50 minutes each) and 3 repetitions for DLS and zeta-potential, respectively, were averaged, with error bars indicating the standard deviation. In both cases the samples were kept at 25 °C during the measurements. The Smoluchowski approximation was the model employed for the zeta-potential measurements.

Fluorescence spectroscopy was performed with a Horiba Fluorolog 3 spectrometer using right-angle illumination equipped with a 450 W Xenon lamp and a FL-1030-UP photomultiplier as a detector. Measurements were carried out with an excitation wavelength of 530 nm (3 nm bandpass) and the emission was collected from 550-800 nm (3 nm bandpass) with a 1 nm increment step.

Cryo-EM measurements were performed by Dr. Davide Demurtas, and EELS and EDX measurements were carried out by Dr. David F. Reyes Vasquez at the CIME-EPFL facilities.

**Surface density quantification.** DBCO-Cy5 was reacted with NP-$N_3$, and the extinction spectra before (mix) and after (NPs) purification were measured to calculate the Cy5 concentration (extinction at 650 nm) normalized by the $SiO_2$ concentration (extinction at 350 nm) in each case. The ratio of Cy5 / NP was obtained and normalized by the surface of a sphere using the following formula:

$$Cy5\ \left[\frac{Cy5}{nm^2}\right] = \frac{E_{650nm}^{NPs} \cdot V^{NPs}}{E_{350nm}^{NPs}} \cdot \frac{E_{350nm}^{mix}}{E_{650nm}^{mix} \cdot V^{mix}} \cdot C_{Cy5}^{mix} \cdot \frac{1}{4\pi r^2} \qquad (1),$$

where $E_{650nm}$ and $E_{350nm}$ are the measured extinction intensities at 650 and 350 nm, respectively, for the NPs both before (mix) and after (NPs) purification, $C_{Cy5}$ is the concentration of Cy5 in the reaction mixture, and r the radius of the NPs in nm. The volume was applied to normalized volume differences before (mix) and after (NPs) purification. The DNA surface density was calculated in an analogous way by measuring the extinction intensity at 260 nm

$$DNA\ \left[\frac{DNA}{nm^2}\right] = \frac{E_{260nm}^{NPs} \cdot V^{NPs}}{E_{350nm}^{NPs}} \cdot \frac{E_{350nm}^{mix}}{E_{260nm}^{mix} \cdot V^{mix}} \cdot C_{DNA}^{mix} \cdot \frac{1}{4\pi r^2} \qquad (2).$$

**DNA origami synthesis.** The DNA origami design employed in our studies was designed using CaDNAno[97]. The structure is available at https://nanobase.org/structure/146[98]. A 7249-nucleotide long scaffold extracted from the M13mp18 bacteriophage (Bayou Biolabs LLC) was folded into the desired shape using 243 staples in 1x TAE (40 mM Tris, 10 mM acetate, 1 mM EDTA), 12 mM $MgCl_2$, pH 8 buffer. It was mixed in a 10-fold excess of staples (purchased from IDT) over scaffold, and 100-fold for the functional staples (fluorophores, biotin, and handles, purchased from Biomers GmbH) shown in **Table S1**. The mixture was heated to 70 °C and cooled to 25 °C at a rate of 1 °C every 20 min. The DNA origami structures were later purified by 1% agarose (LE Agarose, Biozym Scientific GmbH) gel electrophoresis at 70 V for 2 h and stored at 4 °C.[99]

**Self-assemble of NPs to DNA origami.** For the synthesis of the monomer and dimer structures, DNA-functionalized NPs (1:1 ratio of T18:T28) were mixed overnight with the DNA origami in a molar ratio of 1:5 and 1:10 (Origami : NPs) for monomers and dimers, respectively, in 0.75x TAE 9 mM $MgCl_2$ buffer. Samples were purified by gel electrophoresis at 70 V for 3 h using 0.3% agarose containing 0.5x TAE and 6 mM $MgCl_2$. The desired bands were cut out, squeezed, and deposited on custom-made chromium grids coated with BSA-biotin and neutravidin (BSA-biotin and neutravidin, 0.5mg/mL in PBS, were consecutively incubated on the grid for 1hs) via biotin protrusions from the bottom of the origami structures. Single-molecule fluorescence measurements for the visualization of DNA origami structures containing single ATTO 647N fluorophores were carried out on a custom-built widefield microscope[99,100] and later colocalized with the NPs observed on a TESCAN Mira 3 scanning electron microscope.

**Acknowledgements**

The authors would like to thank Prof. Fernando Stefani, Prof. Marco Lattuada, Prof. David J.Pine, Prof. Stefano Sacanna, and Prof. Joonsuk Oh for many fruitful discussions. Moreover, the authors would like to express their gratitude to Dr. Davide Demurtas and Dr. David F. Reyes for their assistance in acquiring, analyzing, and discussing Cryo-EM, and EELS and EDX measurements, respectively.

G.P.A. acknowledges support from the Swiss National Science Foundation (200021_184687) and the National Center of Competence in Research Bio-Inspired Materials NCCR (51NF40_182881). G. P. A. and D. G. thank the European Union Program HORIZON-Pathfinder-Open: 3D-BRICKS, grant Agreement 101099125. D.G. thanks the European Union under the Horizon 2020 Program, FET-Open: DNA-FAIRYLIGHTS, Grant Agreement 964995. H.S. acknowledges the supported by Kobe University Strategic International Collaborative Research Grant.

# Supporting Information

## Universal click-chemistry approach for the DNA functionalization of nanoparticles

*Nicole Siegel*[1]*, Hiroaki Hasebe*[2]*, Germán Chiarelli*[1]*, Denis Garoli*[3,4]*, Hiroshi Sugimoto*[2]*, Minoru Fujii*[2]*, Guillermo P. Acuna*[1,5]**, Karol Kołątaj*[1,5]***

[1] Department of Physics, University of Fribourg, Chemin du Musée 3, Fribourg CH 1700, Switzerland.

[2] Department of Electrical and Electronic Engineering, Graduate School of Engineering, Kobe University, Kobe 657-8501, Japan.

[3] Dipartimento di Scienze e Metodi dell'Ingegneria, Università di Modena e Reggio Emilia, Via Amendola 2 Padiglione Tamburini, 42122 Reggio Emilia, Italy.

[4]Istituto Italiano di Tecnologia, Via Morego 30, 16163, Genova, Italy

[5] Swiss National Center for Competence in Research (NCCR) Bio-inspired Materials, University of Fribourg, Chemin des Verdiers 4, CH-1700 Fribourg, Switzerland.

*Corresponding authors.

Email address: guillermo.acuna@unifr.ch, karol.kolataj@unifr.ch.

**Supplementary Table 1: DNA sequences**. The table lists all DNA staples used in the project that has been modified compared to the original design[98]. Handles are staples extended with 8 adenines that can bind NPs functionalized with the complementary sequences on either the left or the right side of the DNA origami. ATTO 647N staple is an oligonucleotide incorporating a single ATTO 647N molecule in the structure. Biotin staples facilitates binding between the DNA origami and a neutravidin-functionalized glass surface.

| **Name** | **Sequence** |
|---|---|
| Handle Left 1 | TACAAATTGCCAGTAAAGTAATTCTGTCCAGAAAAAAAA |
| Handle Left 2 | TAAAGACTGTTACTTAGGCGCAGACGGTCAATAAAAAAAA |
| Handle Left 3 | GCCACCCTTCGATAGCATAATCCTGATTGTTTAAAAAAAA |
| Handle Left 4 | CACCAACCAAGTACAAGTACAGACCAGGCGCAAAAAAAA |
| Handle Left 5 | CGACATTCCCAGCAAAATTATTTGCACGTAAAAAAAAAAA |
| Handle Left 6 | CCTTTTTTTTCATTTCAACAATAACGGATTCGAAAAAAAA |
| Handle Left 7 | ACGGGTAATAAATTGTTGACCAACTTTGAAAGAAAAAAAA |
| Handle Left 8 | ATTAATTATGAAACAATATACAGTAACAGTACAAAAAAAA |
| Handle Left 9 | AATATTGACGTCACCGTGCGTAGATTTTCAGGAAAAAAAA |
| Handle Left 10 | AACAGTAGCCAACATGACATGTTCAGCTAATGAAAAAAAA |
| Handle Left 11 | AAGAACGCAAGCAAGCATAATATCCCATCCTAAAAAAAAA |
| Handle Left 12 | CTTGCGGGGTATTAAAAAACCAATCAATAATCAAAAAAAA |
| Handle Left 13 | AAATCAGATCATTACCATCAACAATAGATAAGAAAAAAAA |
| Handle Left 14 | CATATGGTAGCAAGGTAATGGAAGGGTTAGAAAAAAAAA |
| Handle Left 16 | CCATCTTTCGTTTTCAAACCACCAGAAGGAGCAAAAAAAA |
| Handle Left 17 | GAGCCACCATCAAGTTTCCTGATTATCAGATGAAAAAAAA |
| Handle Right 1 | GAGCACGTGCAAGTGTGAGAAAGGAAGGGAAGAAAAAAAAAA |
| Handle Right 2 | CATCTGCCACCCGTCGTGAGGAAGGTTATCTAAAAAAAAA |
| Handle Right 3 | ATTATTACTTGGGAAGTTCATTACCCAAATCAAAAAAAAA |
| Handle Right 4 | GTTTGATAAGAACGTAGTTTTTTGGGGTCGAAAAAAAAA |
| Handle Right 5 | ATTTAGGATTAAGAACCATTCAGTGAATAAGGAAAAAAAA |
| Handle Right 6 | TACCCCGGTTAAAATTCCTTTGCCCGAACGTTAAAAAAAA |
| Handle Right 7 | GTAATCGTGCTCATTTCTTTACAAACAATTCGAAAAAAAA |
| Handle Right 8 | CACGCTGGAAACCGTCATGGCCCACTACGTGAAAAAAAAA |
| Handle Right 9 | ACGTTGGTGGATTGACGGTCAGTTGGCAAATCAAAAAAAA |
| Handle Right 10 | ACATAACGACTTTAATACACCAGAACGAGTAAAAAAAAA |
| Handle Right 11 | CCTTATAATTGTTCCAAAATCGGAACCCTAAAAAAAAAAA |
| Handle Right 12 | GCTACAGCACACCCGGAGCTTGACGGGGAAAAAAAAAA |
| Handle Right 13 | GCTATTAGTTAACACCCAAATGAAAAATCTAAAAAAAAAA |
| Handle Right 14 | CTAAAACACAGAAGATAACCTCAAATATCAAAAAAAAAAA |
| Handle Right 15 | GAGTCTGGAAATAATTGATAATACATTTGAGGAAAAAAAA |
| Handle Right 16 | CGGCCTCACTTTCATCACTAACAACTAATAGAAAAAAAAA |
| Atto647N staple | CCCC TACCGACAAAAGGTAATAAGAGAATATAAAG TT Atto647N |
| Biotin staple 1 | ACAGGAAGATTGTCCCCCTTATTCACCCTCATTTGTTTC-biotin |
| Biotin staple 2 | GTTGATAGATATAAGCATAAGTATAGC-biotin |
| Biotin staple 3 | AGAGTACTCACGCTAACCTTTAATTGC-biotin |
| Biotin staple 4 | CACTAAAACACTCACGAACTAACACTAAAGT-biotin |
| Biotin staple 5 | TCACGACGTTGGGCGCTTTGGTAAAAC-biotin |
| Biotin staple 6 | CAGAGATAGCGATAGTGAATAACATAA-biotin |

**Suplementary Figures**

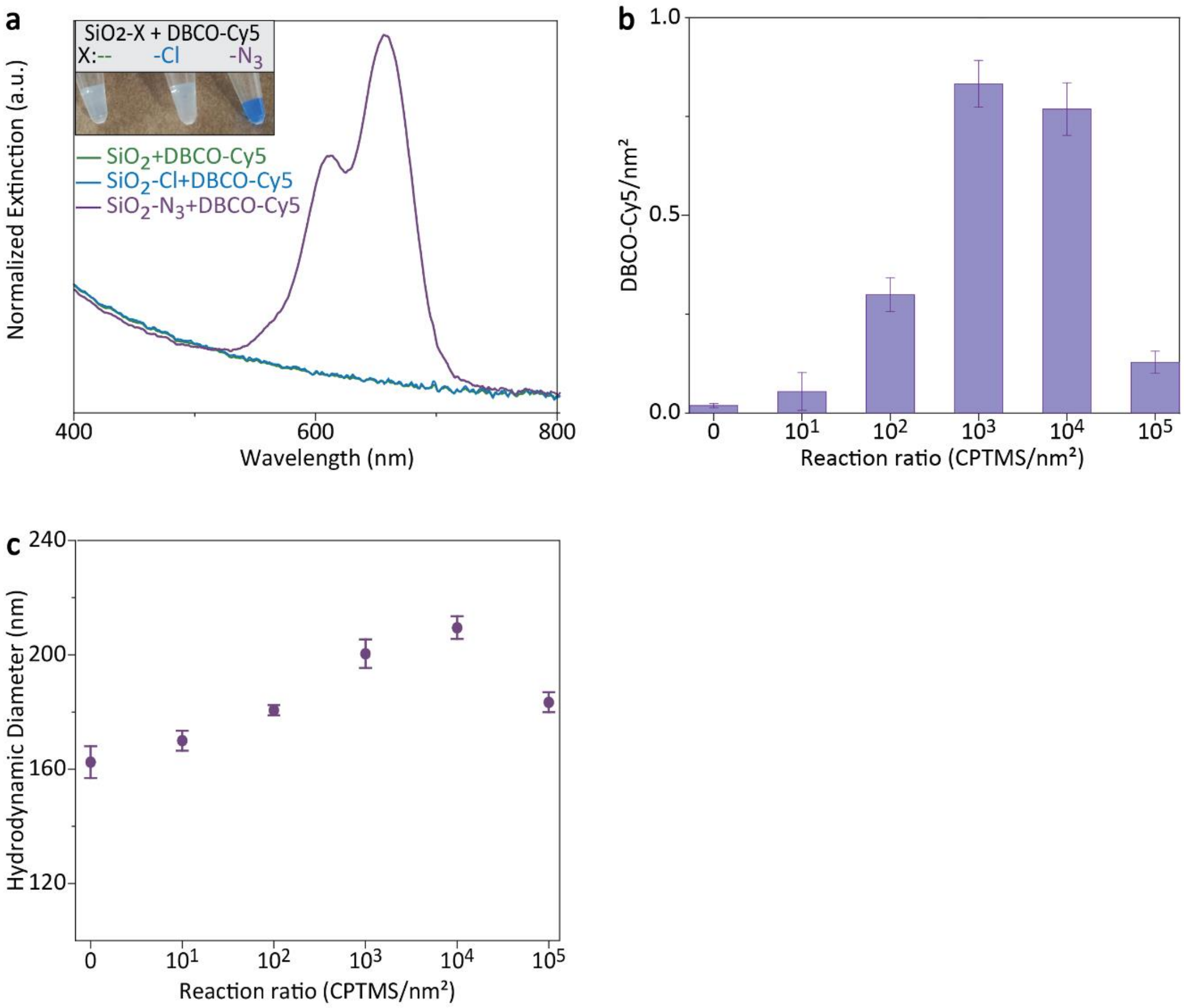

**Figure S1: Azide functionalization of $SiO_2$ NPs.** (a) Normalized extinction spectra of $SiO_2$ (green) $SiO_2$-Cl (blue) and $SiO_2$-$N_3$ (violet) NPs after reaction with DBCO-Cy5. $SiO_2$ and $SiO_2$-Cl curves overlap showing no significant difference. Inset: image showing each solution after reaction with DBCO-Cy5 and washed to remove any free dye. (b) Cy5 surface density of $SiO_2$ NPs functionalized under different CPTMS/$nm^2$ reaction ratios calculated by extinction at 650 nm (i.e., Cy5 absorbance maximum) according to equation 1 (**Materials and Methods)**. (c) Corresponding hydrodynamic diameter measurements.

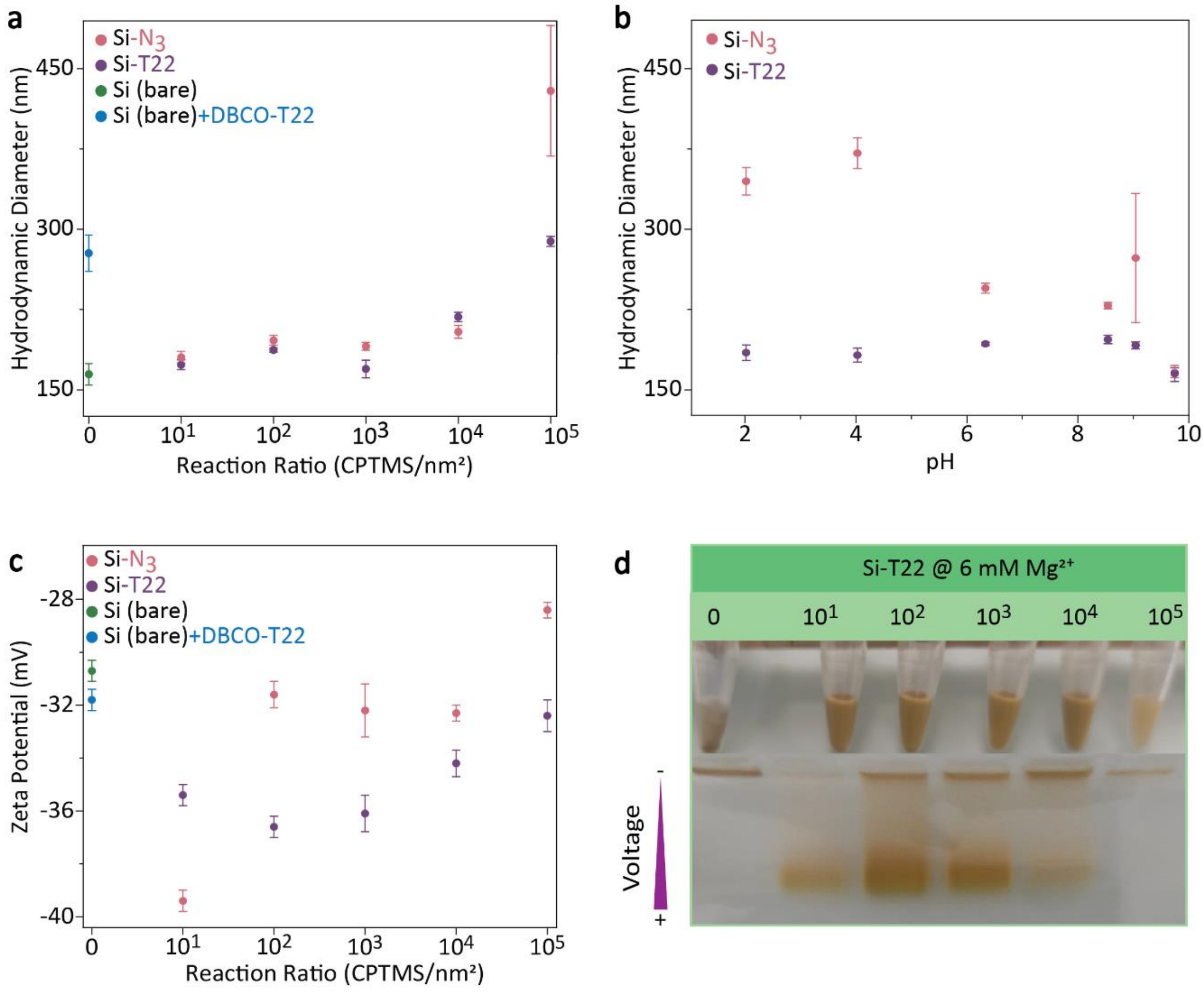

**Figure S2: Functionalization of Si NPs.** (a) Hydrodynamic diameter of Si NPs functionalized with different CPTMS/nm$^2$ reaction ratios before, $N_3$ (pink), and after, T22 (violet), freezing-assisted SPAAC with DBCO-T22; bare Si (i.e., particles without CPTMS) before (blue) and after freezing (green) are shown as reference. (b) Hydrodynamic diameter of Si NPs functionalized with $N_3$ (pink) and T22 (violet) using a reaction ratio of $10^3$ CPMTS/nm$^2$ at different pH values. (c) Zeta potential of $N_3$ (pink) and T22 (violet) for different CPTMS/nm$^2$ reaction ratios with the corresponding bare Si NPs controls (green and blue respectively). (d) Image and agarose gel electrophoresis of Si-T22 for different CPTMS/nm$^2$ reaction ratios.

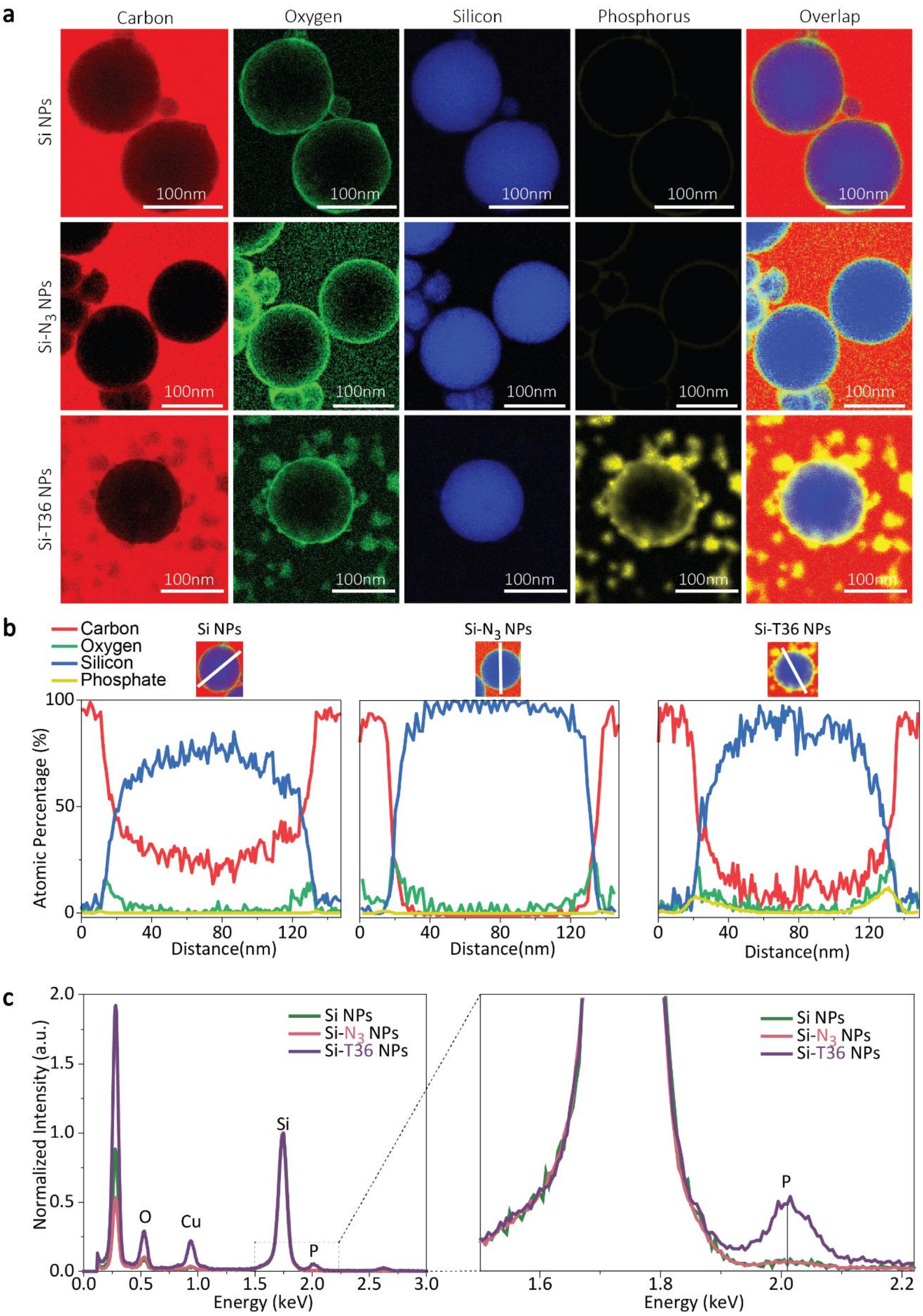

**Figure S3: Elemental Analysis.** (a) EELS elemental analysis of a Si NP ($d$ = 140 nm) with non-bearing functionalization (upper row), functionalized with azide (middle row) and T36 DNA strands (lower row) showing (from left to right) carbon signal (red), oxygen signal (green), silicon signal (blue), phosphorus

signal (yellow), and their overlap. (b) Atomic percentage of the three samples (from left to right: bare, azide and DNA functionalized NPs) along the shown lines. (c) Complementary EDX analysis of bare (green), azide (pink) and DNA (violet) functionalized Si NPs with a magnification of the phosphorus pick on the right.

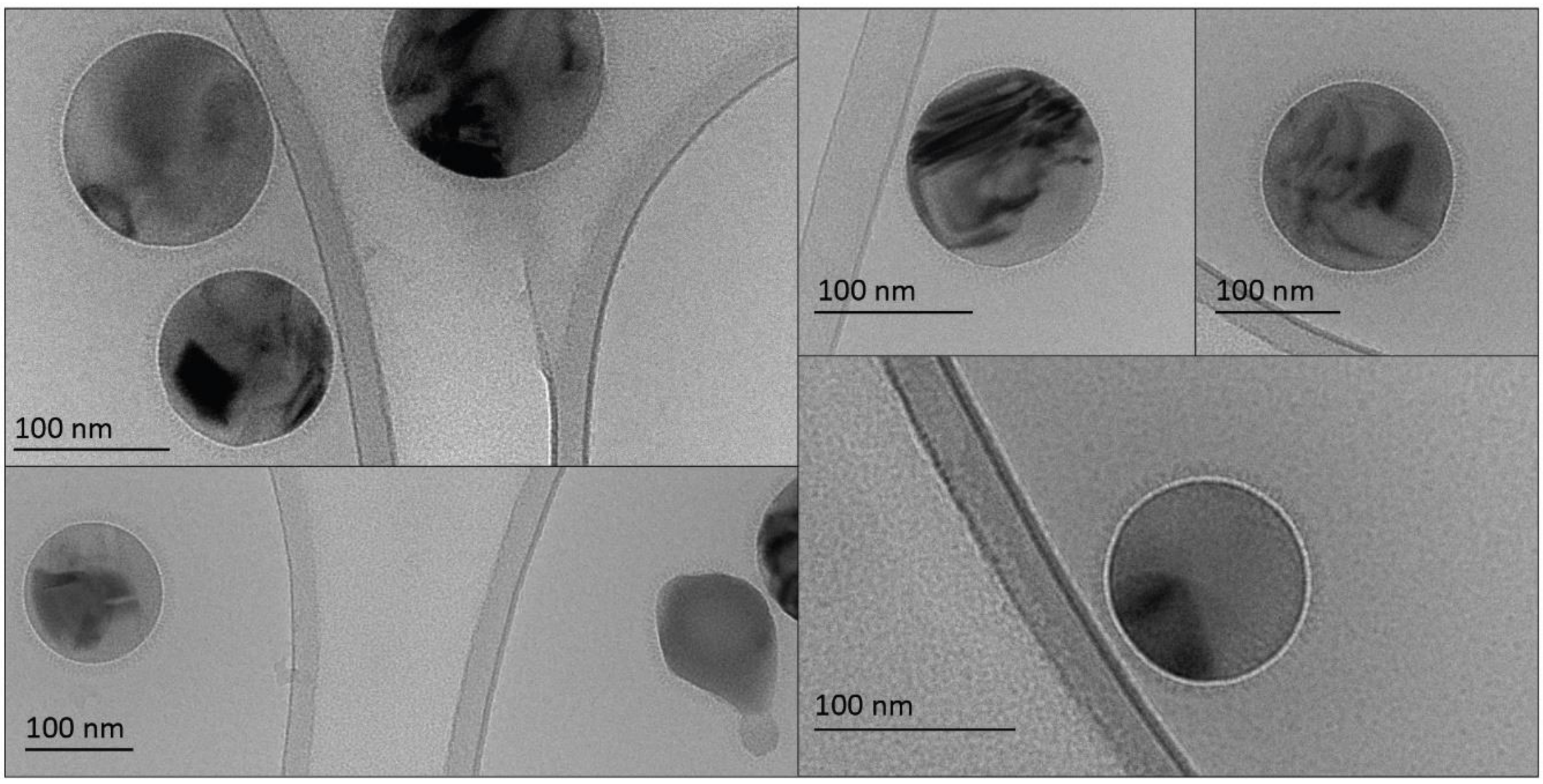

**Figure S4: Imaging of Si-DNA NPs.** Cryo-EM images of T36 DNA-functionalized Si NP ($d$ = 140 nm).

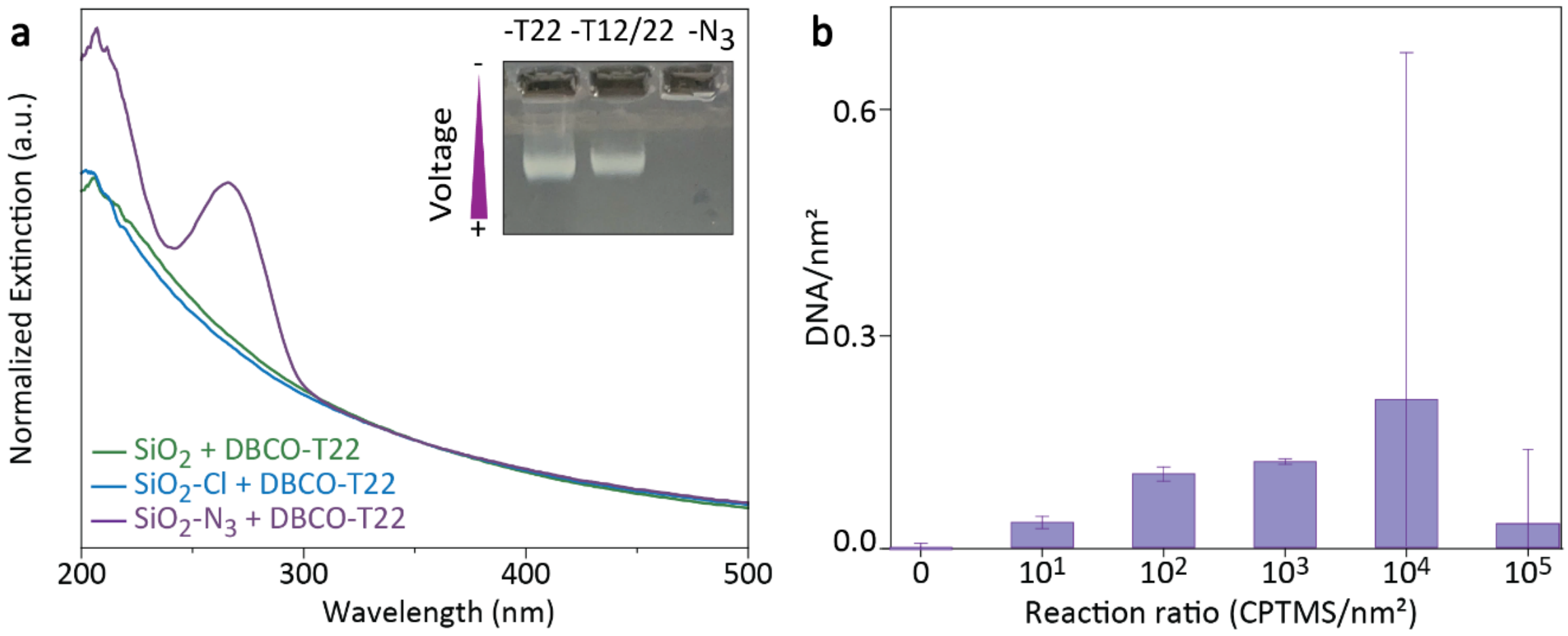

**Figure S5: DNA functionalization of $SiO_2$ NPs.** (a) Normalized extinction spectra of $SiO_2$ (green) $SiO_2$-Cl (blue) and $SiO_2$-$N_3$ (violet) NPs after reaction with DBCO-T22. Inset: image showing the result of a gel electrophoresis run carried out in PBS, SDS at 6mM $Mg^{2+}$ for the DNA (either solely T22, or a 50:50 mixture of T12 and T22) and azide functionalized $SiO_2$ NPs. (b) DNA surface density of $SiO_2$ NPs functionalized under different CPTMS/ $nm^2$ reaction ratios; calculated by extinction at 260 nm (i.e., DNA absorbance maximum) according to equation 2 (**Materials and Methods)**.

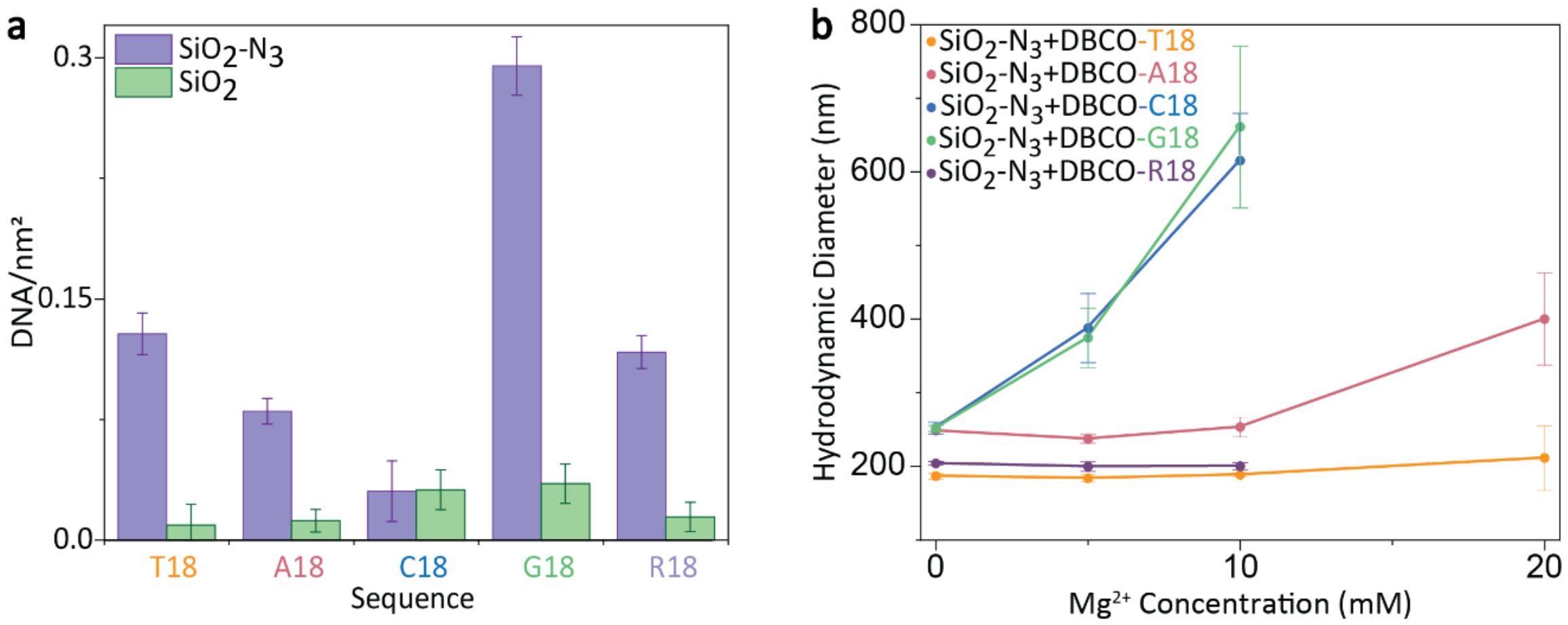

**Figure S6: Sequence dependent DNA functionalization of $SiO_2$ NPs.** (a) DNA surface density (calculated by extinction at 260 nm according to equation 2, see **Materials and Methods**) of $SiO_2$ (green) and $SiO_2$-$N_3$ (violet) after freezing-assisted SPAAC with DBCO-T18 (orange), DBCO-A18 (pink), DBCO -C18(blue), DBCO -G18 (green), DBCO -R18 (violet). (b) Hydrodynamic diameter of $SiO_2$ NPs as a function of salt concentration. $SiO_2$ NPs were functionalized with either $N_3$ + DBCO-T18 (orange), $N_3$ + DBCO-A18 (pink), $N_3$ + DBCO-C18 (blue), $N_3$ + DBCO-G18 (green) and $N_3$ + DBCO-R18 (violet). No data is included for G18-$SiO_2$ and C18-$SiO_2$ NPs above 15mM $Mg^{2+}$ due to the fact that these particles formed extremely large aggregates. Error bars represent one standard deviation, and solid lines serve as a guide to the eye.

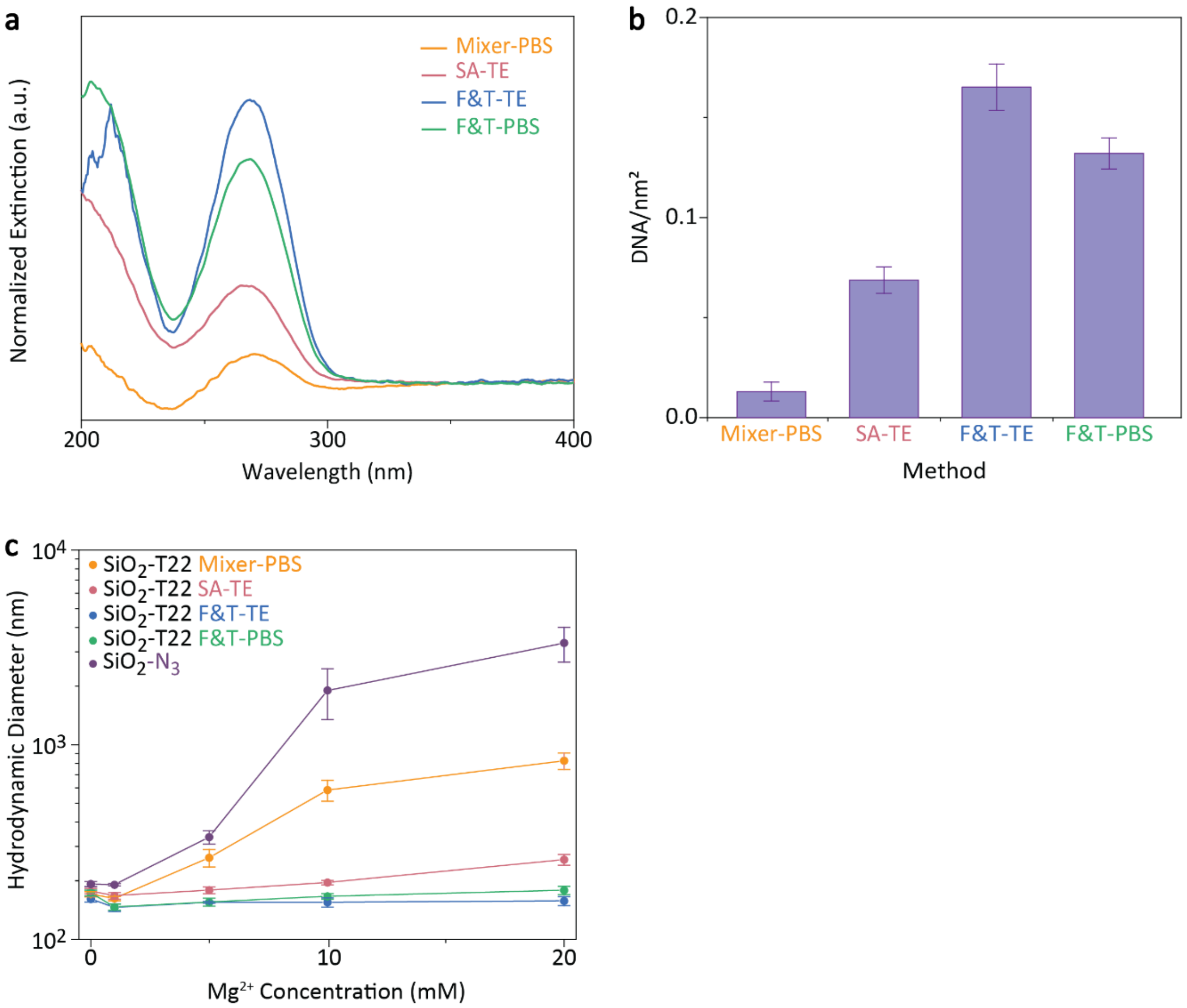

**Figure S7: Method dependent DNA functionalization of $SiO_2$ NPs.** (a) Normalized extinction spectra of $SiO_2$ NPs after functionalization with DBCO-T22 using room-temperature mixing for 48 hours in phosphate-buffered saline (Mixer-PBS, orange), salt aging in Tris-EDTA buffer (SA-TE, pink), or freezing-assisted SPAAC in Tris-EDTA buffer (F&T-TE, blue) or phosphate-buffered saline (F&T-PBS, green) and (b) corresponding surface density calculated by extinction at 260nm according to equation 2 (**Materials and Methods**). (c) Hydrodynamic diameter of $SiO_2$ NPs as a function of salt concentration. $SiO_2$ NPs were functionalized with either $N_3$ (violet) or DBCO-T18 using room-temperature mixing for 48 hours in phosphate-buffered saline (Mixer-PBS, orange), salt aging in Tris-EDTA buffer (SA-TE, pink), or freezing-assisted SPAAC in Tris-EDTA buffer (F&T-TE, blue) or phosphate-buffered saline (F&T-PBS, green). Error bars represent one standard deviation, and solid lines serve as a guide to the eye.

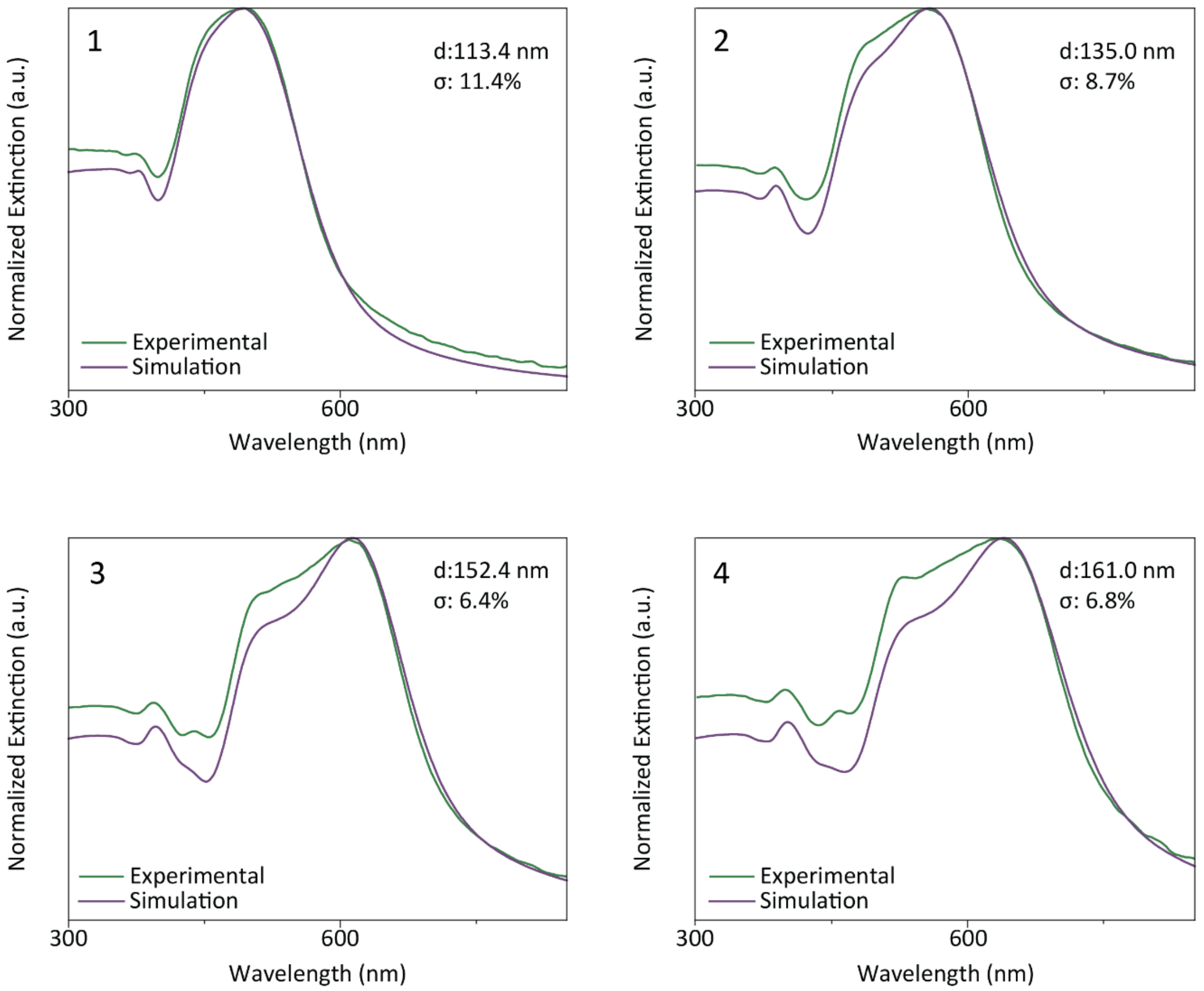

**Figure S8: Size separation of Si NPs.** Measured (green) and simulated (violet) extinction spectra for size-separated Si-T22 NPs, indicating the average particle diameter (d) and standard deviation (σ) as calculated from the simulated spectra. Panels 1-4 correspond to the numbered aliquots of size-separated Si-T22 particles retrieved from gel electrophoresis as shown in **Figure 3f**. Spectra were simulated by applying the Mie Theory for spherical Si NPs.

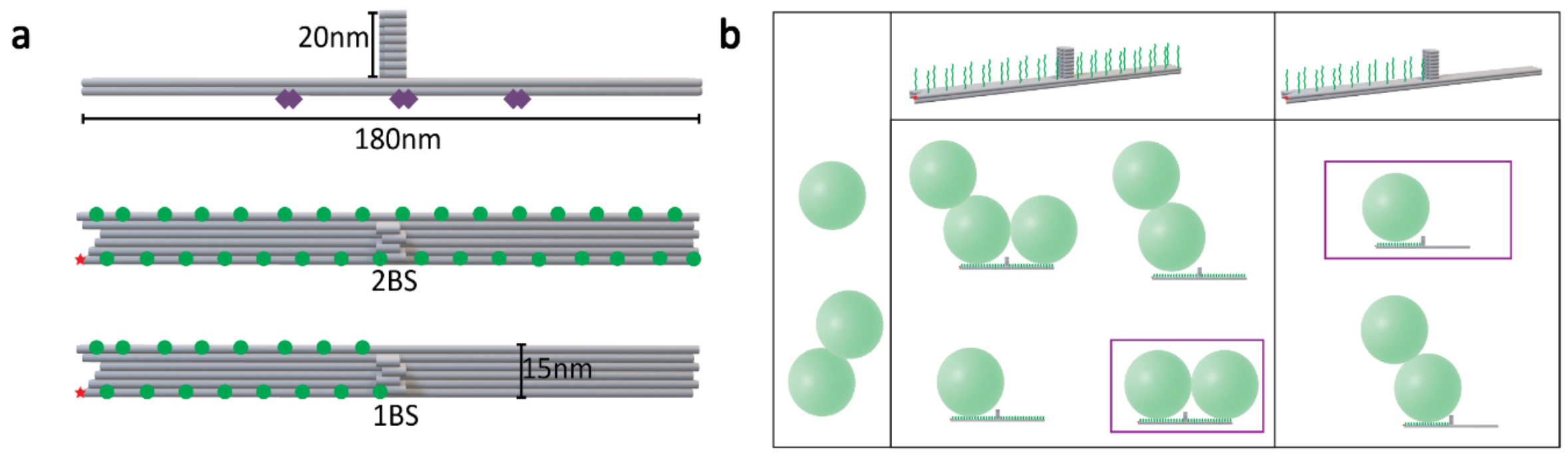

**Figure S9: Assembly with DNA origami.** (a) Schematic depicting the DNA-origami structures. The DNA origami contains two rows of poly-adenine (green dots) on one or both sides of the mast for the formation of Si monomers (1BS) or dimers (2BS), a single ATTO 647N fluorophore (red star), and six biotin molecules (violet diamonds) for the binding to the surface of a glass slide. (b) Possible structures obtained after mixing 2BS or 1BS structure with Si NPs with preexisting dimerization. The envision structures are boxed.

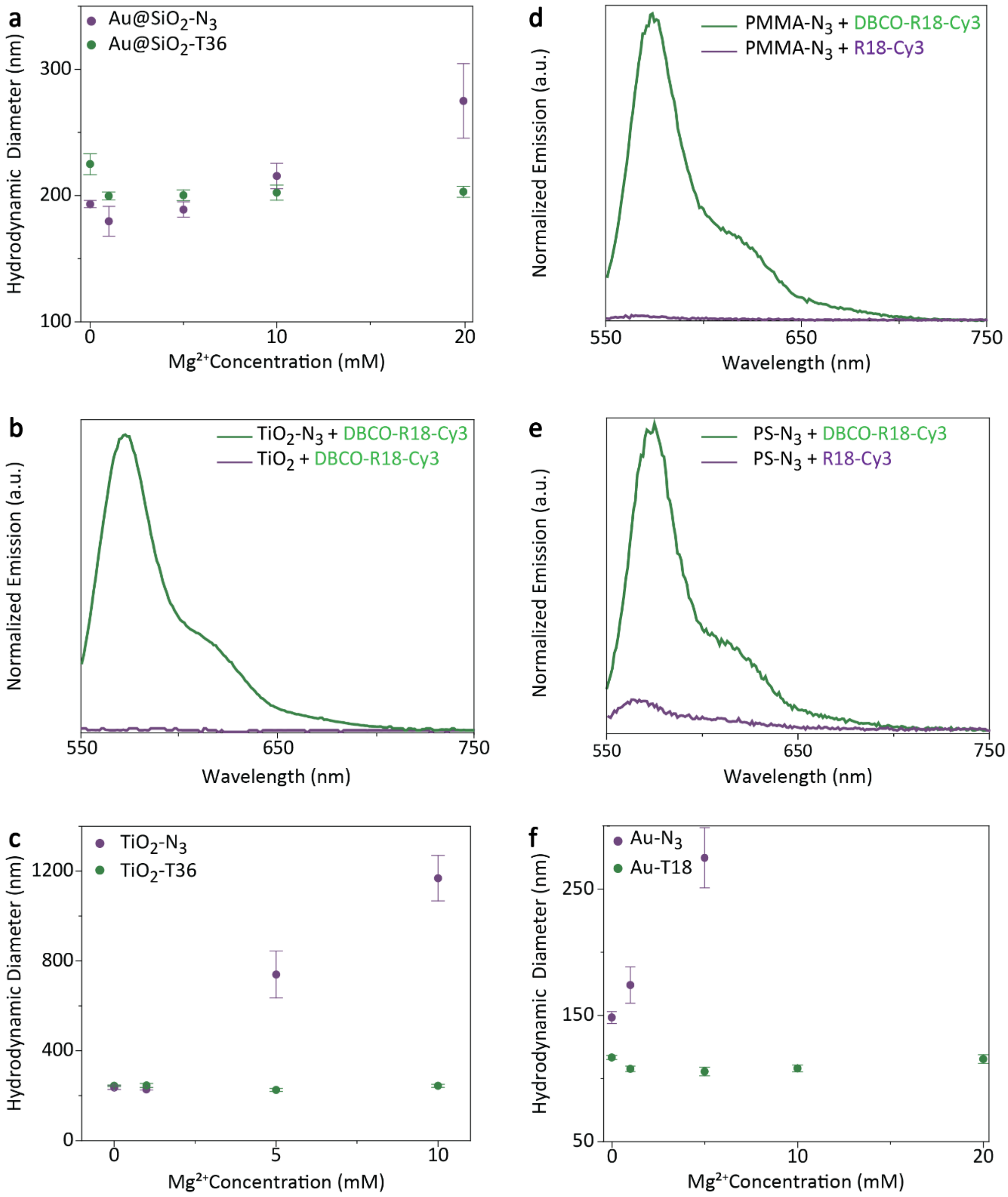

**Figure S10: Versatility of freezing-assisted SPAAC.** (a) Hydrodynamic diameter of Au@$SiO_2$ as function of salt concentration before ($N_3$) and after (T36) DNA functionalization. (b) Normalized emission spectra of $TiO_2$ NPs with non-bearing functionalization ($TiO_2$, violet) or functionalized with azide ($TiO_2$-$N_3$, green) after reaction with DBCO-R18-Cy3. (c) Hydrodynamic diameter of $TiO_2$ NPs functionalized with azide ($TiO_2$-$N_3$, violet) or T36 DNA strands ($TiO_2$-$N_3$, violet) as function of salt concentration. Normalized emission spectra of azide-functionalized (d) PMMA and (e) PS NPs after reaction with DBCO-R18-Cy3 (green) or R18-Cy3 (violet). (f) Hydrodynamic diameter of Au NPs functionalized with azide (Au-$N_3$, violet) or T18 DNA strands (Au-$N_3$, violet) as function of salt concentration. No data is included for Au-$N_3$ NPs above 5mM $Mg^{2+}$ due to the fact that these particles formed extremely large aggregates.